%% file: ms.tex
\documentclass{emulateapj}
\usepackage{apjfonts}

\usepackage{ifpdf}
\usepackage{graphicx}
\usepackage{natbib}

\ifpdf\DeclareGraphicsExtensions{.pdf,.jpg,.tif}\else\DeclareGraphicsExtensions{.ps,.eps}\fi

\newcommand \microjy{$\mu$Jy}

\newcommand \dog{IRBG}
\newcommand \dogs{IRBGs}
\newcommand{\Ion}[2]{#1\,{\small #2}}
\newcommand \spitzer{\textit{Spitzer}}
\newcommand \chandra{\textit{Chandra}}
\newcommand{\Msun}{M_{\odot}}

\newcommand{\Lsun}{L_{\odot}}

\def\gsim{\mathrel{\rlap{\lower4pt\hbox{\hskip1pt$\sim$}} \raise1pt\hbox{$>$}}}
\def\lsim{\mathrel{\rlap{\lower4pt\hbox{\hskip1pt$\sim$}} \raise1pt\hbox{$<$}}}

\slugcomment{Accepted for publication in The Astrophysical Journal}
\shorttitle{IR-Bright/Optically-Faint Galaxies}
\shortauthors{DONLEY ET~AL}

\begin{document}

\title{The AGN, Star-forming, and Morphological Properties of Luminous IR-Bright/Optically-Faint Galaxies}

\author{J. L. Donley, \altaffilmark{1,2} 
	G. H. Rieke, \altaffilmark{2}
	D. M. Alexander, \altaffilmark{3}
	E. Egami, \altaffilmark{2}
	P. G. P\'{e}rez-Gonz\'{a}lez \altaffilmark{4,5}}

\altaffiltext{1}{Space Telescope Science Institute, 3700 San Martin Drive, Baltimore, MD 21218; Giacconi Fellow; donley@stsci.edu}
\altaffiltext{2}{Steward Observatory, University of Arizona, 933 
North Cherry Avenue, Tucson, AZ 85721}
\altaffiltext{3}{Department of Physics, Durham University, Durham DH1 3LE, UK}
\altaffiltext{4}{Departamento de Astrof\'{\i}sica y CC. de la Atm\'osfera, Facultad de
CC. F\'{\i}sicas, Universidad Complutense de Madrid, 28040 Madrid, Spain}
\altaffiltext{5}{Associate Astronomer at Steward Observatory, The University of Arizona}

\begin{abstract}

We present the AGN, star-forming, and morphological properties of a sample of 13 MIR-luminous ($f_{\rm 24} \gsim 700$~\microjy) IR-bright/optically-faint galaxies (IRBGs, $f_{\rm 24}/f_{\rm R} \gsim 1000$).  While these $z \sim 2$ sources were drawn from deep \chandra\ fields with $>200$ ks X-ray coverage, only 7 are formally detected in the X-ray and four lack X-ray emission at even the $2 \sigma$ level.   \spitzer\ IRS spectra, however, confirm that all of the sources are AGN-dominated in the mid-IR, although half have detectable PAH emission responsible for $\sim 25\%$ of their mid-infrared flux density.  When combined with other samples, this indicates that at least 30-40\% of luminous \dogs\ have star-formation rates in the ULIRG range ($\sim100-2000 \Msun$~yr$^{-1}$).   X-ray hardness ratios and MIR to X-ray luminosity ratios indicate that all members of the sample contain heavily X-ray obscured AGN, 80\% of which are candidates to be Compton-thick. Furthermore, the mean X-ray luminosity of the sample, log~$L_{\rm 2-10 keV}$(ergs~s$^{-1}$)$\sim44.6$, indicates that these \dogs\ are Type 2 QSOs, at least from the X-ray perspective.  While those sources most heavily obscured in the X-ray are also those most likely to display strong silicate absorption in the mid-IR,  silicate absorption does not always accompany X-ray obscuration.  Finally, $\sim 70\%$ of the \dogs\ are merger candidates, a rate consistent with that of sub-mm galaxies (SMGs), although SMGs appear to be physically larger than \dogs.  These characteristics are consistent with the proposal that these objects represent a later, AGN-dominated, and more relaxed evolutionary stage following soon after the star-formation-dominated one represented by the SMGs.

\end{abstract}


\section{Introduction}
\input{table1}

Studies of heavily-obscured star-forming galaxies and active galactic nuclei (AGN) in the distant Universe have lagged behind those of their unobscured counterparts, due largely to their being extremely faint in the optical and UV. This optically faintness, however, need not be a limitation and can instead be used as a selection criterion, as dust-enshrouded sources faint at short wavelengths should be comparably bright in the infrared where their absorbed radiation is re-emitted.  One might therefore expect sources with bright infrared emission yet faint optical magnitudes ($f_{\rm 24}/f_{\rm R} \gsim 1000$) to be ideal obscured galaxy/AGN candidates, and indeed, such sources comprised some of the first \spitzer\ Space Telescope \citep{werner04} targets.

Initial studies of the brightest ($f_{\rm 24} \gsim 700$~\microjy) IR-bright/optically-faint galaxies (referred to here as IRBGs) targeted for \spitzer\ IRS mid-infrared spectroscopic \citep{houck04} follow-up observations indicated that most lie at $z\sim2$ and have either featureless spectra or spectra dominated by silicate absorption, properties indicative of AGN activity \citep[e.g.,][]{houck05,yan05,weedman06red}.    Their optical faintness was therefore attributed both to distance as well as to obscuration by the dust surrounding an AGN's central engine, although more recent studies suggest that the AGN's host galaxy may also contribute significantly to the observed extinction \citep{brand07,polletta08}.  

Subsequent multiwavelength studies have since confirmed the AGN nature of the most luminous \dogs, although they reach differing conclusions concerning the relative importance of star-formation and AGN activity amongst fainter ($f_{\rm 24} < 700$~\microjy) \dogs\ \citep[e.g.,][]{dey08,donley08,georgantopoulos08,fiore08,pope08dog,desai09,treister09,fiore09}. Furthermore, they have shown that the space density of the most luminous \dogs\ is comparable to that of luminous, unobscured AGN at $z=2$ \citep{dey08}.  If these sources are in fact heavily obscured AGN, they therefore represent an important phase in the growth of supermassive black holes during the era in which both star-formation and AGN activity peaked. 

To constrain the level of obscuration in these luminous AGN, studies turned to the X-ray.  Unfortunately, the large-area survey fields from which the luminous, and therefore rare, \dogs\ were initially selected (e.g. the Spitzer First-Look Survey, the NOAO Deep Wide Field Survey, and the SWIRE Survey) and then followed-up with IRS have minimal X-ray coverage capable of detecting only the brightest AGN \citep{dey08}.   Nonetheless, studies in these fields suggest that at least $\sim 50\%$ and perhaps as many as 95\% of \textit{X-ray--detected} \dogs\ with $f_{\rm 24} > 1.3$~mJy are X-ray--obscured AGN \citep{lanzuisi09}, some of which may be Compton-thick \citep{polletta06,alexander08ct}.

Alternatively, other studies have focused on samples of \dogs\ selected in fields with deeper X-ray coverage, such as the \chandra\ Deep Fields and COSMOS.  The deepest X-ray fields, however, are also the smallest, so those studies most suited to constrain X-ray obscuration are also those in which the focus generally shifts to the fainter and more numerous population of \dogs\ whose nature remains controversial.  In the larger COSMOS field, however, \citet{fiore09} find that 40\% of \dogs\ with moderate flux densities of $f_{\rm 24} > 550$~\microjy\ are detected in the X-ray with a mean rest-frame obscured luminosity of log~L$_{\rm {2-10 keV}} $(ergs~s$^{-1}$)$= 43.5$ and an X-ray hardness ratio (HR=(H-S)/(H+S), where H=1.5-6 keV and S=0.3-1.5 keV) of 0.50, consistent with obscured yet Compton-thin absorption.  The stacked X-ray signal from the remaining 60\% of the sample has a similar HR of 0.53, although its interpretation is dependent on a number of factors including the assumed underlying column density distribution, the intrinsic MIR/X-ray luminosity ratio, and most importantly, the assumed photon index of emission from star formation.  Assuming a star-forming X-ray photon index ($\Gamma$) of 1.9, \citet{fiore09} conclude that 94\% of the X-ray non-detected, $f_{\rm 24} > 550$~\microjy\ \dogs\ are heavily obscured, and likely Compton-thick, AGN.  If a harder X-ray photon index of $\Gamma \sim 1.0-1.4$ (as observed in the starburst-dominated ULIRGS of \citet{franceschini03}, \citet{ptak03}, and \citet{teng05}) is assumed, however, the obscured AGN fraction of this moderate-luminosity sample falls significantly.

\input{table2}
\input{table3}

To bridge the gap between these infrared and X-ray studies and therefore better constrain the luminosity, obscuration, and power sources of these cosmologically interesting AGN, we have obtained IRS spectra of a sample of \dogs\ selected in deep X-ray fields with effective exposures of $T_{\rm x} > 200$~ks.  Because we do not require that our targets be \textit{detected} in the X-ray, however, we do not bias our sample towards the brightest or least-obscured galaxies \citep[e.g.,][]{brand08}.  The sample discussed below is therefore the first uniformly-selected sample of \dogs\ with both deep X-ray and IRS coverage.   Because they were selected from deep multiwavelength fields, many of the sources also have \textit{Hubble} Space Telescope (HST) ACS imaging, enabling a morphological study of this unbiased sample and a comparison to sub-mm galaxies (SMGs), a population of $z\sim2$ star-forming galaxies proposed by some to be the merger-induced progenitors of luminous \dogs\ \citep[e.g.,][]{dey08,pope08dog,coppin10,desika09smg,desika09dog,brodwin08}. 

The paper is organized as follows.  In \S2, we discuss the sample selection, observations, and data reduction.  The optical and infrared photometric properties of the sources are then examined in \S3.   In \S4, we present the IRS spectra and discuss the AGN and star-forming contribution to the MIR emission of these sources.  The X-ray emission is discussed in \S5, as is the agreement between the X-ray and IR properties, and the star-formation rates of the sources are discussed in \S6.  In \S7 we summarize the morphological properties of the \dogs' hosts and compare them to those of the SMGs.  The discussion follows in \S8, and we then summarize our conclusions in \S9.   Throughout the paper, we assume the following cosmology:($\Omega_{\rm m}$,$\Omega_{\rm \Lambda},H_0$)=(0.27, 0.73, 70.5~km~s$^{-1}$~Mpc$^{-1}$), and quote all magnitudes in the AB system unless otherwise noted.

\section{Sample Selection and Observations}

To ensure deep X-ray coverage, we selected the \dog\ sample from the \chandra\ Deep Fields North and South \citep[CDF-N, CDF-S, $T_{\rm x} \sim 2$~Ms;][]{alexander03,luo08}, the Extended \chandra\ Deep Field South \citep[E-CDFS, $T_{\rm x} \sim 250$~ks;][]{lehmer05}, and the Extended Groth Strip \citep[EGS, $T_{\rm x} \sim 200$~ks;][]{laird09}, the combined area of which is 0.96 deg$^{2}$.  We chose for IRS follow-up sources with extreme IR/optical flux ratios ($f_{\rm 24}/f_{\rm R} \gsim 1000$) typical of those in \cite{houck05}, \cite{yan05}, and \cite{weedman06red,weedman06irs} based on the \spitzer\ MIPS 24 \micron\ and optical photometry compiled in \cite{pgperez08} and the UCM Extragalactic Database\footnote{http://guaix.fis.ucm.es/$\sim$pgperez/Proyectos/ucmcsdatabase.en.html}.  In addition, we required flux densities in excess of $f_{\rm 24} \gsim 700$~\microjy\ to guarantee that IRS spectra could be obtained in a reasonable amount of time.  The final sample is comprised of 13 \dogs.

The 3 sources in the CDF-N were observed with IRS as part of programs 20456 (PI Chary) and 20733 (PI Urry), and the spectra of two (IRBG5 and IRBG7) can be found in \citet{pope08smg} and \citet{murphy09}.  For the remaining 10 sources, we obtained 1st and 2nd order spectra with the IRS Long-Low (LL) module (program 30419, PI Rieke).  Observation details are given in Table 1.  The resulting wavelength range of 14-37 \micron\ guarantees coverage of the 9.7~\micron\ silicate absorption feature at $z = 0.4-2.8$ and the 7.7~\micron\ aromatic (hereafter polycyclic aromatic hydrocarbon (PAH)) feature at $z = 0.8-3.8$.  Of the 13 sources in our sample, 4 have optical/NIR spectroscopic redshifts from the literature, placing them at $1.6\le z \le 2.0$ \citep{szokoly04,swinbank04,davis07}.

\subsection{Data Reduction}
The IRS data were reduced using the IrsLow package developed by D. Fadda to accurately measure low-resolution spectra of faint, high redshift sources \citep[see][for more details]{fadda10}.  Briefly, this package corrects for residual background, rogue pixels (pixels with high dark current and/or photon responsivity), and cosmic rays, and takes into account all frames produced by the IRS/SSC pipeline.  The background and noise images are produced by masking the target spectrum (and any serendipitous spectra) on each frame and then coadding the resulting frames.  A biweight statistical estimator is then used iteratively and interactively to minimize the contamination from deviant pixels, and pixels that deviate by more than $5 \sigma$ from the mean local value are flagged as rogue.  After rejecting any cosmic ray events and taking the spectral distortion into account, the spectra are then weighted by the PSF, optimally extracted, and coadded to produce the final spectrum.

\section{Optical/IR Photometric Properties}

We list in Table 2 the optical and IR characteristics of the \dogs.  The 13 sources in our sample have 24 \micron\ flux densities ranging from 701 to 2299~\microjy\ (median $f_{\rm 24} = 1069$~\microjy), R-band magnitudes ranging from 23.7 to 25.3 AB (median $R=24.8$ AB), and 24 \micron\ to R-band flux ratios ranging from 1204 to 3961 (median $f_{\rm 24}/f_{\rm R} = 2152$).

\begin{figure*}
\epsscale{1.1}
\plotone{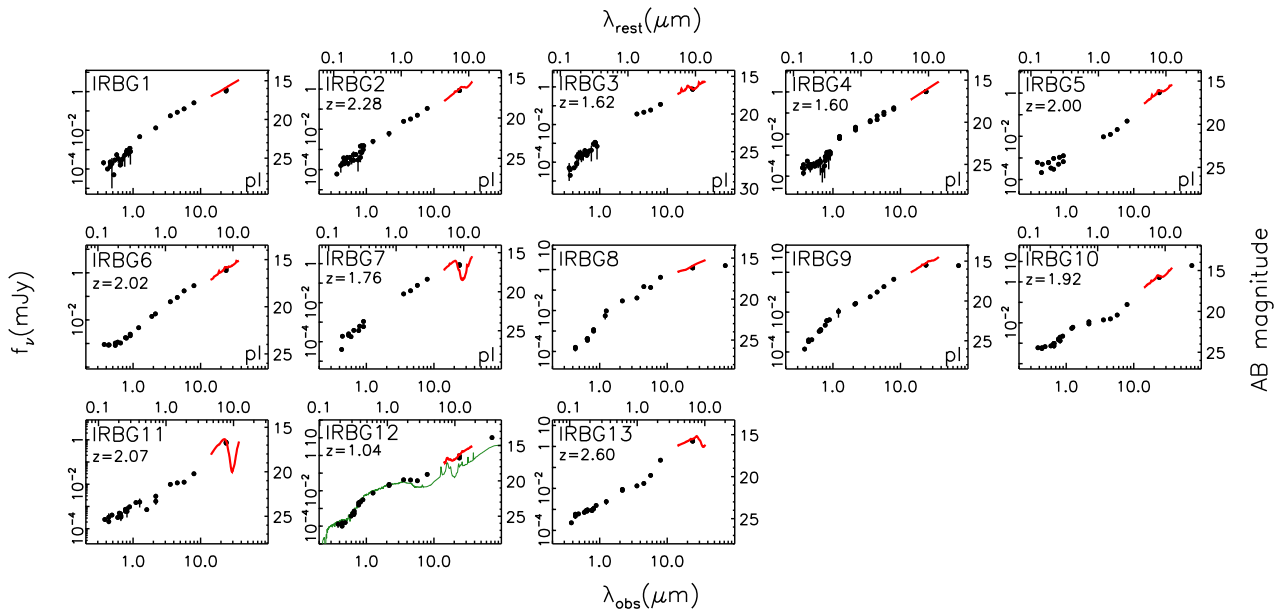}
\caption{Broad-band SEDs of the 13 \dogs.  The majority show power-law SEDs that extend from the near to mid-IR.  Those sources that meet the formal IRAC power-law criterion (see \S3.1) are indicated by a 'pl' in the lower right corner. The best fits to the IRS spectra (see \S4) are shown in solid (red) lines and the template of NGC 6090 is over-plotted (in green) on the SED of IRBG12 to confirm its low redshift ($z\sim1$; see \S4.1).}
\end{figure*}

To place these sources in the context of recent studies, we also show in Table 3 the IR-excess selection criteria met by each of the 13 \dogs.  All sources meet the DOG (dust-obscured galaxy) criteria of \citet[$f_{\rm 24}/f_R \ge 1000$, $f_{\rm {24}}>0.3$mJy]{dey08}, and all but 2 meet one or more of the \citet[$f_{\rm {24}}>0.75$mJy, $R_{\rm Vega} > 24.5$]{houck05}, \citet[log$(\nu f_\nu({\rm 24 \micron})/\nu f_\nu({\rm 8 \micron})) \ge 0.5$, log$(\nu f_\nu({\rm 24 \micron})/\nu f_\nu({\rm 0.7 \micron})) \ge 1.0$]{yan05}, or \citet[$f_{\rm {24}}>1.0$mJy, $R_{\rm Vega} > 23.9$]{weedman06red,weedman06irs} selection criteria.  Furthermore, nine sources meet the IR-excess criteria of \citet[$f_{\rm 24}/f_R \ge 1000, (R-K)_{\rm Vega} \ge 4.5$]{fiore08} or \citet[$f_{\rm 24}/f_R \ge 1000, R_{\rm AB} - m_{\rm 3.6\micron, AB} > 3.7$]{georgantopoulos08}, and eight meet the combined MIPS/IRAC criteria defined by \cite{polletta08} to select highly luminous and heavily obscured AGN.

\vspace*{1cm}
\subsection{SEDs}

The broad-band (0.4-70\micron) SEDs of the \dogs\ are shown in Figure 1.  We constructed the SEDs by first compiling aperture-matched photometric catalogs using the CDF-S data of \citet[][$RIz$]{marzke99}, \citet[][$JK$]{vandame01}, \citet[][$UU_{\rm p}BVRI$]{arnouts02}, COMBO17 \citep{wolf04}, \citet[][$bvizJHK$]{giavalisco04}, \citet[][$I$]{lefevre04}, and GALEX ($FUV,NUV$), the CDF-N data of \citet[][\textit{bviz}]{giavalisco04} and \citet[][\textit{UBVRIz'HK'}]{capak04}, and the EGS data ($UBgVRIzJK$) of \citet{villar08} and Barro et al. (2010), in preparation.  For more details on the available datasets and the aperture-matching procedure, see \cite{pgperez05,pgperez08} and the UCM Extragalactic Database (see footnote 6).

While a handful of the source SEDs shown in Figure 1 display a weak stellar bump (the broad 1.6 \micron\ feature that dominates the SEDs of star-forming galaxies and which tends to be correlated with the presence of PAH emission in \dogs\ \citep[e.g.][]{desai09}), the majority of the sources in our sample have power-law SEDs indicative of AGN activity.  When we apply the IRAC (3.6-8.0 \micron) power-law selection criteria used by \cite{aah06} and \cite{donley07} to identify AGN-dominated sources, we find that 7 of the 13 sources are indeed power-law AGN and one additional source (IRBG5) meets the criteria if we lower the required chi-squared probability from $P{\chi} > 0.1$ to $P{\chi} > 0.01$ \citep[e.g.,][]{aah06}\footnote{For consistency with the IRAC power-law selection in \cite{aah06} and \cite{donley07}, we assume an IRAC flux calibration uncertainty of 10\%.}. Because the IRAC photometry of IRBG8 is questionable due to its position near the edge of the IRAC field, the total power-law fraction is $\sim8/12$, or $\sim 70\%$.  Our results are therefore in agreement with those of \cite{dey08}, who find that while only $\sim 20\%$ of \dogs\ at $f_{\rm 24} = 300$~\microjy\ have power-law SEDs, the power-law fraction rises to $\sim 55\%$ at $f_{\rm 24} = 700$~\microjy\ (our flux cut) and to $\sim 70-85\%$ at $f_{\rm 24} > 1000$~\microjy\ (the median flux density of our sample). Furthermore, all of the power-law sources in our sample are extremely red, with a mean slope of $\alpha = -1.99$ (where $f_{\rm \nu} \propto \nu^{\alpha}$), and are therefore strong AGN candidates regardless of redshift \citep[see][]{donley07,donley08}.

\subsection{Infrared colors}

A number of infrared color-criteria have been used in the literature to discriminate between AGN and star-formation activity \citep[for a review, see][]{donley08}. While the reliability and completeness of pure color/color cuts is heavily dependent on a source's redshift, \cite{pope08dog} find that for \dogs\ (which typically lie at $z \sim 2$), $f_{\rm 8.0 \micron}/f_{4.5 \micron} = 2.0$ provides a convenient dividing line between star-formation--dominated ($f_{\rm 8.0 \micron}/f_{4.5 \micron} < 2.0$) and AGN-dominated ($f_{\rm 8.0 \micron}/f_{4.5 \micron} > 2.0$) emission \citep[also see][who suggest an even lower cut of 1.65]{coppin10}.  As a confirmation of this method, we plot in Figure 2 the expected colors of a number of AGN and star-forming galaxies.  We find that only AGN fall redward of the cut, as expected, although we note that lower-luminosity AGN (e.g., Seyfert galaxies) as well as self-absorbed AGN \citep{treister09} could display bluer colors more typical of star-forming galaxies.  As shown in Figure 2, however, all of our \dogs\ lie in the red AGN-dominated regime, with 3 sources showing a significantly higher $f_{\rm 24 \micron}/f_{8.0 \micron}$ ratio than the rest.  \cite{pope08dog} attribute this excess of 24 \micron\ emission either to intrinsic obscuration indicative of AGN activity, which enhances the observed-frame 24 \micron\ flux and/or obscures the observed-frame 8 \micron\ flux, or to the 7.7 \micron\ PAH feature passing into the MIPS 24\micron\ band.  We explore these possibilities further in \S4.3.

\begin{figure}
\vspace*{0.5cm}
\epsscale{1.1}
\plotone{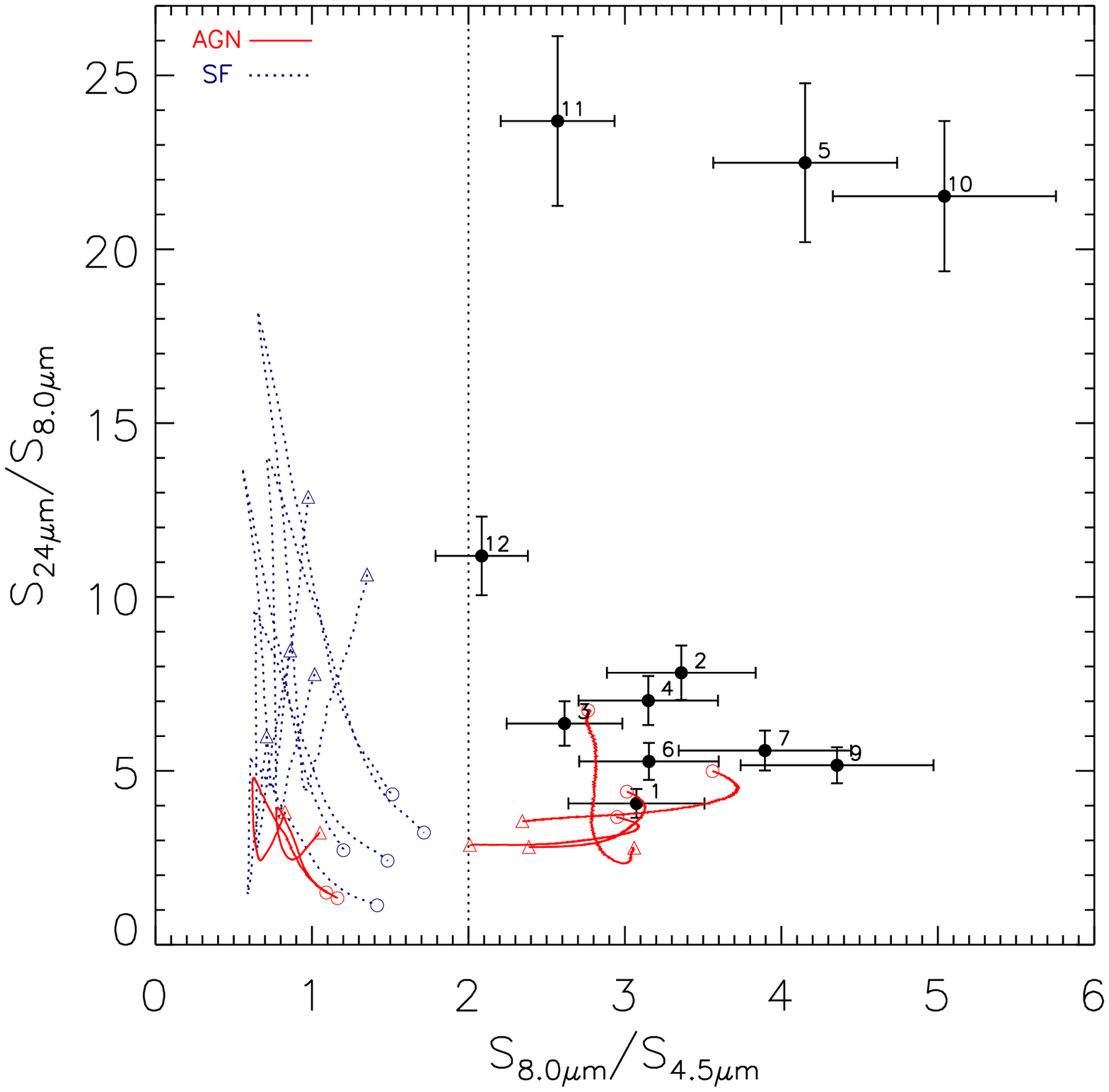}
\caption{MIR colors of the \dogs, where the numbers give the source ID.  Overplotted are the redshifted ($z=1-3$) templates of AGN from \citet[TQSO, Seyfert 2, Seyfert 1.8, Mrk231]{polletta08} and \citet[AGN and AGN2]{assef08} and of purely star-forming ULIRGS from \cite{rieke09}.  The two AGN templates with colors blueward of $S_{\rm 8.0}/S_{\rm 4.5} = 2$ are the Seyfert 1.8 and Seyfert 2 templates of \cite{polletta08}.  Triangles give the template colors at $z=1$ and circles given the colors at $z=3$.  All of the \dogs\ have $S_{\rm 8.0}/S_{\rm 4.5} > 2$ as expected only for AGN-dominated objects \citep[e.g.,][]{pope08dog}.  IRBG13, with its rapidly rising IRAC continuum, falls off the plot at values of $S_{\rm 8.0}/S_{\rm 4.5} = 12.8$ and $S_{\rm 24}/S_{\rm 8.0} = 8.1$. }
\end{figure}

\begin{figure*}
\epsscale{1.0}
\plotone{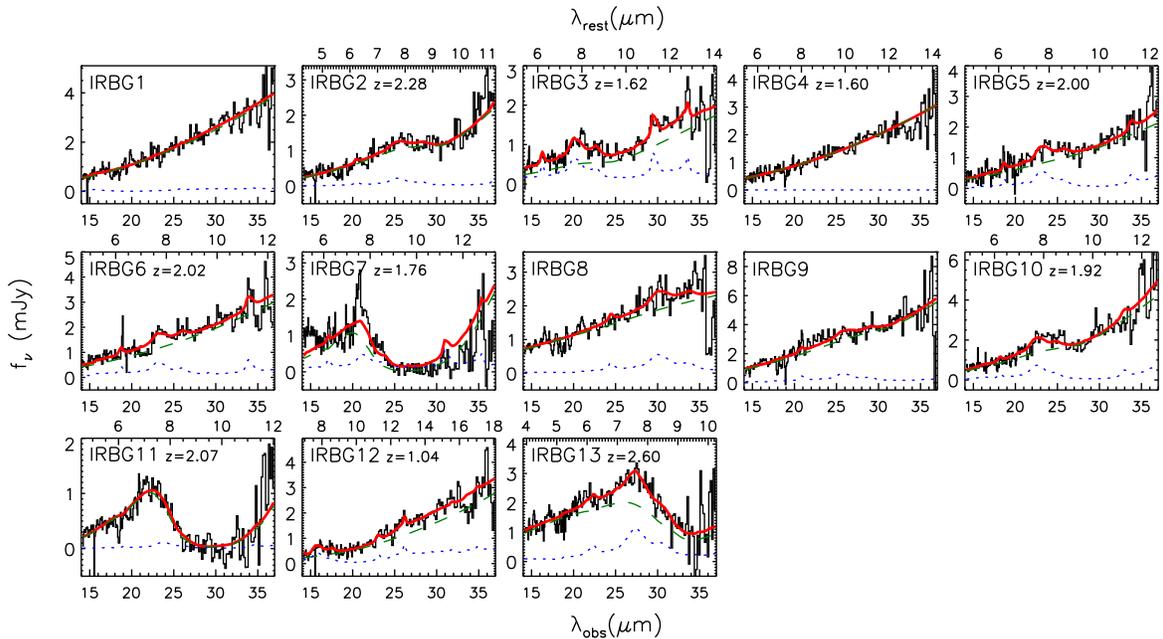}
\caption{IRS spectra of the \dogs.  The green dashed line shows the observed contribution from the AGN whereas the blue dotted line shows the observed contribution from star-formation.  The overall fit is given by the red solid line.}
\end{figure*}

\section{IR Spectra}

The IRS spectra of our 13 \dogs\ are shown in Figure 3.  To measure redshifts and determine the source of the emission (e.g., AGN or star-formation), we simultaneously fit a power-law continuum and a star-forming template to each spectrum.  For the star-forming templates, we adopt the starburst, luminous infrared galaxy (LIRG), and ultraluminous infrared galaxy (ULIRG) templates of \cite{rieke09}.  While we allow the power-law component to be reddened according to the extinction law of \cite{draine03}, in no case do we require additional reddening to be applied to the star-forming templates. 

Using a chi-squared minimization routine, we then fit the spectrum while leaving the template normalizations, power-law index, power-law extinction, and redshift as free parameters.  To protect against local minima, we varied the assigned starting values for the parameters, and chose the input values that produced the fit with the lowest reduced chi-squared (if more than one minimum was found).  Finally, to place errors on the fit parameters, we created 1000 simulated spectra for each source (based on the measured spectra and their associated Gaussian errors) and re-ran our fitting program on each.  The errors quoted below for the redshifts (\S4.2) and AGN fractions (\S4.3) are taken to be the standard deviations in the resulting distributions of fit values.  This procedure produces reasonable spectral fits for all sources except IRBG7, discussed below, although for 4 objects the lack of spectral features prevented direct redshift constraints.

\subsection{IRBG7}

IRBG7 has two spectral features and/or artifacts that impact the fit. The first is a sharp emission feature at $\lambda_{\rm obs} \sim 21$~\micron, or $\lambda_{\rm emit}\sim 7.6 \micron$ assuming our best-fit redshift of $z=1.76$ (for comparison, \citealt{murphy09} measure a redshift of $z=1.75$).  This feature, which can be seen in the raw spectral data, could be strong 7.65~\micron\ [\Ion{Ne}{VI}] emission that is partially blended with the 7.7 \micron\ PAH emission. 

The second strange feature in this spectrum is the apparently blue continuum at $\lambda \lsim 18$~\micron, present in both our reduction of the IRS data as well as that of \citet[A. Pope 2009, private communication]{pope08smg} and \cite{murphy09}.  If the AGN continuum fit is free to vary over all spectral indices, it therefore settles on a blue slope of $\alpha=1.69$ (where $f_{\rm \nu} \propto \nu^{\alpha}$), despite its extremely red ($\alpha=-2.21$) IRAC power-law continuum.  Such behavior, however, appears to be unphysical, for while the IRS slope need not match the IRAC slope, the two are generally close. In fact, the offset between the IRAC spectral slope ($\alpha_{\rm IRAC}$) and the IRS spectral slope ($\alpha_{\rm IRS}$) for the 7 additional power-law galaxies in our sample ranges from 2\% to 42\%, with a mean (median) value of only 18\% (16\%).

We therefore adopt a revised method to fit the spectrum of IRBG7.  First, we place an upper limit on the IRS AGN continuum slope by adjusting the measured IRAC slope of $\alpha=-2.21$ upwards (bluewards) by 18\%, the mean offset between the IRAC power-law and IRS slopes.  With this assumed AGN continuum of $\alpha=-1.81$, we then perform an AGN-only fit at $\lambda > 18$~\micron\ to determine the AGN extinction that best fits the profile of the 9.7 \micron\ silicate absorption feature.   We then set the normalization of the AGN continuum by extrapolating the IRAC continuum fit from 8 to 14 \micron.  When we do the same for the other power-law galaxies in the sample, we find that the mean offset between the predicted and observed 14 \micron\ flux density is 35\%. To place a lower limit on the AGN contribution, we therefore scale the AGN contribution downwards until the 14 \micron\ AGN flux density is 35\% lower than predicted by the IRAC extrapolation, while allowing the star-forming contribution to vary so as to give the lowest reduced chi-squared.  The resulting fit under-predicts both the short-wavelength emission and the 21 \micron\ emission peak, as expected, but now places a robust lower-limit on the AGN's contribution to the MIR emission.

\subsection{Redshifts}

Of the 13 sources in our sample, 4 have optical/NIR spectroscopic redshifts from the literature \citep{szokoly04,swinbank04,davis07}.  Using the IRS spectra, we determine reliable redshifts for an additional 6 sources, for an overall redshift completeness of $\sim 80\%$.  The IRS redshifts derived for 3 of the 4 sources with prior optical/NIR redshifts agree to within $\Delta z \le 0.1$ in all cases (The fourth source, IRBG4, has a power-law IRS spectrum from which no redshift estimate can be made).  

Of the 3 remaining sources, two (IRBG1 and IRBG9) have power-law spectra and SEDs, and one (IRBG8) has both a spectrum dominated by power-law emission as well as questionable IRAC photometry due to its position near the edge of the IRAC field, thus preventing accurate spectroscopic or photometric redshift determination.

The mean and median redshifts of our sample, $z=1.89$ and $z=2.00$, respectively, agree quite well with the typical redshifts of similar samples.  For instance, both \citet{weedman06red} and \citet{houck05} find median redshifts of $z=2.1$, \citet{yan05} finds a median redshift of $z=2.3$, and \citet{polletta08} finds a median redshift of $z=2.2$.  

Finally, with the exception of IRBG12, all of the sources lie at $z>1.60$.  The redshift of IRBG12, $z=1.04$, is therefore quite low in comparison both to our and other \citep[e.g.,][]{weedman06red,houck05,yan05,polletta08} samples of luminous \dogs. This source's low redshift, however, appears robust.  The reduced chi-squared fit has 3 minima, one at $z = 0.11$, one at $z = 1.04$, and one at $z=3.19$.  As shown in Figure 1, however, this source has a weak 1.6 \micron\ stellar bump that is well-fit at optical-NIR rest-frame wavelengths by the $z \sim 1$ template of the starburst galaxy NGC~6090 from \citet{polletta06}.  We can therefore rule out the low and high redshift solutions from the IRS spectroscopy  on the basis of the broad-band SED.

\subsection{Source of power: AGN activity or star-formation?}

The fraction of the overall IRS (14-37 \micron) dust-obscured flux density arising from the AGN power-law component (e.g., the AGN fraction), the best-fit power-law indices, and the optical depths at 9.7 \micron\ (for the power-law component) are shown in Table 2.  On the basis of their MIR (14-37 \micron) emission, all of the \dogs\ in our sample are AGN-dominated, with AGN fractions ranging from 71\% to 100\%.  Six of the \dogs\ are nearly pure power-laws with AGN fractions exceeding 90\%, three have hints of PAH emission with AGN fractions between 80\% and 90\%, and the remaining four have noticeable contributions from PAH features with AGN fractions $< 80\%$.  For comparison, only 2 of the 13 SMGs from \citet{pope08smg} and \citet{murphy09} have similarly-derived AGN fractions exceeding 60\%.  One of these 2 sources is IRBG5 from our sample, and the other is an \dog\ with a 24\micron\ flux density that fell below our flux cut of 700 \microjy\ ($f_{\rm 24} =  303$\microjy). 

As discussed above, the mean offset between the IRAC and IRS spectral slopes of the power-law galaxies is only 18\%.  Furthermore, none of the power-law AGN have IRAC and IRS slopes that differ by more than $\sim 40\%$, indicating that the power-law emission seen at rest-frame wavelengths of $\sim 1-3$~\micron\ continues up to $\sim 12$~\micron, and possibly beyond, assuming a typical redshift of $z \sim 2$.

Recall, however, that 3 of the \dogs\ in our sample (IRBG5, IRBG10, and IRBG11), have far-higher ratios of 24 \micron\ to 8 \micron\ flux than the rest ($S_{\rm 24}/S_{\rm 8.0} \gsim 20$, see Figure 2), due either to extremely high levels of extinction extending into the NIR (at $z=2$, the 2.7 \micron\ rest-frame emission is redshifted into the observed 8 \micron\ band) or to a contribution from the 7.7 \micron\ PAH feature.  The IRS spectra suggest that both of the above scenarios contribute to the enhanced 24 \micron\ flux densities of IRBG5 and IRBG10.  At redshifts of $z=2.00$ and $z=1.92$, the 7.7 \micron\ PAH feature, clearly visible in each of the spectra, serves to enhance the observed MIPS flux.  However, star-formation is responsible for only $\sim 25\%$ of the 24 \micron\ flux density in these two sources.  Removal of the star-forming flux therefore lowers the observed 24 \micron\ to 8 \micron\ flux ratios, but only to values of 15.9 and 17.6, still well above the ratios seen for the remaining sources.  The power-law continuum at an observed wavelength of 8 \micron\ is therefore likely to be suppressed, although shorter-wavelength spectral data would be needed to verify this hypothesis. 

The remaining source, IRBG11, differs from the others in that the AGN accounts for 92\% of the MIR flux density, and no noticeable contribution from PAH emission is seen in the spectrum.  In this case, the large 24 \micron\ to 8 \micron\ flux ratio most likely results from extreme MIR obscuration.  Indeed, this source shows one of the largest 9.7 \micron\ opacities in our sample, $\tau_{\rm {9.7}} = 5.0$, corresponding to an $A_{\rm V}$ of 92.5 if we assume the conversion of \cite{draine03}.  

\input{table4}

\section{X-ray Properties}

As discussed in the introduction, the study of luminous \dogs\ has suffered thus far from a relative lack of sensitive X-ray coverage in the fields where these sources have traditionally been identified (e.g. the \textit{Spitzer} First-Look Survey, the NOAO Deep Wide Field Survey, and the SWIRE survey).  As such, the X-ray analysis of \dogs\ has lagged behind their study in the infrared, and to date, no studies have obtained both IRS spectra \textit{and} deep X-ray imaging of a statistically significant sample of \dogs\ selected independently of their X-ray properties.  The deep X-ray coverage and complete IRS sampling of our uniformly-selected \dogs\ will therefore help to bridge the gap between prior X-ray and MIR spectral studies of these sources.

\subsection{X-ray Detection Fraction}

The X-ray data for the CDF-N, CDF-S, E-CDFS-S, and EGS come from \cite{alexander03}, \cite{luo08}, \cite{lehmer05}, and \cite{laird09}, respectively.  At the position of the 13 \dogs, the X-ray exposures range from $\sim 200$~ks to $\sim 2$~Ms, with a mean value of $\sim 800$~ks.  Only 7 of the 13 \dogs\ ($\sim 50\%$), however, are formally detected in the X-ray (e.g., are detected to high enough significance to be included in published X-ray catalogs).  Their X-ray fluxes in the full (0.5-8 keV), hard (2-8 keV), and soft (0.5-2 keV) X-ray bands are given in Table 4.  For sources in the EGS, we scale the published 0.5-10~keV and 2-10~keV fluxes to the bands defined above assuming an observed photon index of $\Gamma = 1.4$ \citep[the observed photon index of the typical obscured AGN, and thus of the X-ray background, in this energy range, e.g.,][]{marshall80,gendreau95,hickox06}.

For the 6 sources not detected in the catalogs listed above, we searched for faint X-ray counterparts using the procedure outlined in \cite{donley05}.  Briefly, we first calculated the appropriate 60\%, 70\%, and 80\% encircled energy radii (EER) for each source by exposure-weighting the EER, as measured using the lookup tables of \cite{laird09}, for each individual \chandra\ observation. After measuring the source counts in each of the three apertures, we measured sky counts in 10,000 randomly-placed apertures of the same size that (1) did not intersect the source aperture,  that (2) lie within 1\arcmin\ of the source position, and that (3) did not contain a known X-ray source.  In addition, we corrected the counts in each sky aperture to reflect the exposure time measured for the source, and only allowed sky apertures in which the exposure differs from the source exposure by no more than 15\%.  We then fit a Poisson distribution to the resulting distribution of sky counts to determine the significance of any source detection, and chose the EER that maximized the source signal.

Of the 6 formally X-ray non-detected sources, IRBG3 is detected in both the full and hard bands to 4.8 and 5.5 $\sigma$ above the sky, respectively, and IRBG9 is marginally detected in the full band, but only to $2.1 \sigma$ above the sky. The remaining 4 sources remain undetected in all bands, although 1 (IRBG5) lies too close to a known X-ray source to test for very faint X-ray emission.   We place conservative $3 \sigma$ upper limits on the flux of each X-ray--undetected source by adding any positive source counts to a $3\sigma$ upper limit on the measured sky background in the 70\% EER, assuming the typical observed X-ray photon index of an obscured AGN, $\Gamma = 1.4$.

\subsection{X-ray Obscuration}

For the 8 \dogs\ with a formal or weak detection in the hard and/or soft X-ray bands, we crudely estimate the intrinsic column density using the hard to soft X-ray flux ratios, the measured redshifts (or $z=2$ for sources lacking a secure redshift measurement), and an assumed intrinsic (e.g., unabsorbed) X-ray photon index of $\Gamma = 2.0$ \cite[e.g.,][]{george00}.  This method, commonly used to estimate the column densities of faint X-ray sources lacking sufficient counts for spectral fitting, is necessarily approximate in nature as it assumes an X-ray spectrum represented by a single absorbed power-law and neglects the potential effects of more complicated spectral components such as the soft X-ray excess and/or Compton reflection.  Nonetheless, column densities estimated in this way are broadly consistent with those determined via X-ray spectral fitting \citep[see, for example, Appendix B of][]{perola04}.  For instance, the column density we estimate for IRBG6, log~$N_{\rm H} $(cm$^{-2}$)=  23.6, is in good agreement with the spectrally-derived value of \cite{alexander05}, log~$N_{\rm H} $(cm$^{-2}$)$=23.8$. 

The column density estimates are given in Table 4.  To place measurement errors on the columns, we take into account both the errors in the hard and soft X-ray flux, as well as the uncertainty on the redshift.  For those sources lacking a secure redshift measurement, the redshift error is taken to be the standard deviation in the redshift distribution for the full sample of \dogs, $\sigma_{\rm z} = 0.42$. Of the 8 sources for which X-ray--based estimates could be made, 7 are formally obscured with log~$N_{\rm H} $~(cm$^{-2}$)$ \ge 22$ and 1 has an upper limit consistent with significant obscuration (log~$N_{\rm H}$~(cm$^{-2}$)$ \lsim {22.8}$).

From their infrared properties, we know that {\it all} of our sample contain strong AGN.  To place crude column density estimates on the X-ray--non-detected \dogs, and to obtain independent estimates for the X-ray--detected sources, we follow \cite{alexander08ct} and \cite{lanzuisi09} and explore the observed relationship between an AGN's rest-frame X-ray and MIR luminosity \citep[e.g.,][]{lutz04}. The position of our \dogs\ in MIR/X-ray color space is shown in Figure 4, where the rest-frame 6 \micron\ luminosities ($\nu L_{\rm \nu}$) have been estimated using the best-fit AGN contribution to the IRS spectra (which directly sample the rest-frame 6 \micron\ emission at $z> 1.4$) and the rest-frame 2-10 keV luminosities have been estimated using the observed soft band flux (0.5-2 keV at $z=0$, 1.5-6 keV at $z=2$) and an assumed observed photon index of $\Gamma = 1.4$, typical of obscured AGN. The shaded region in Figure 4 gives the intrinsic observed relation for local AGN \citep{lutz04} and the solid line gives the intrinsic luminosity-dependent relation of \cite{maiolino07}, who drew their AGN sample from both low ($z<0.2$) and high ($z=2-3$) redshifts\footnote{We assume the  average IRS spectral slope of our sample, $\alpha = -1.94$, to convert the \cite{maiolino07} relation, initially defined at 6.7 \micron, to 6.0 \micron.}.   As shown, the average \textit{absorption-corrected} (e.g. intrinsic) luminosity ratio of the 6 sources in our sample with crude X-ray--derived column density estimates is consistent with the \cite{maiolino07} relation.  The effect of various obscuring columns on the observed relation of \cite{maiolino07} is shown by dashed lines, where we again assume an intrinsic photon index of $\Gamma = 2.0$.  We also assume for simplicity that the 6 \micron\ emission from the torus is isotropic, i.e. that it does not depend on the obscuring column or on the angle at which the AGN/torus is viewed.  If the torus is optically thick at MIR wavelengths \citep[e.g.][]{pier92,heckman95,buchanan06}, however, an obscured AGN would fall to both lower X-ray \textit{and} lower MIR fluxes, thus increasing somewhat the column density estimated via this method.  

\begin{figure}
\epsscale{1.1}
\plotone{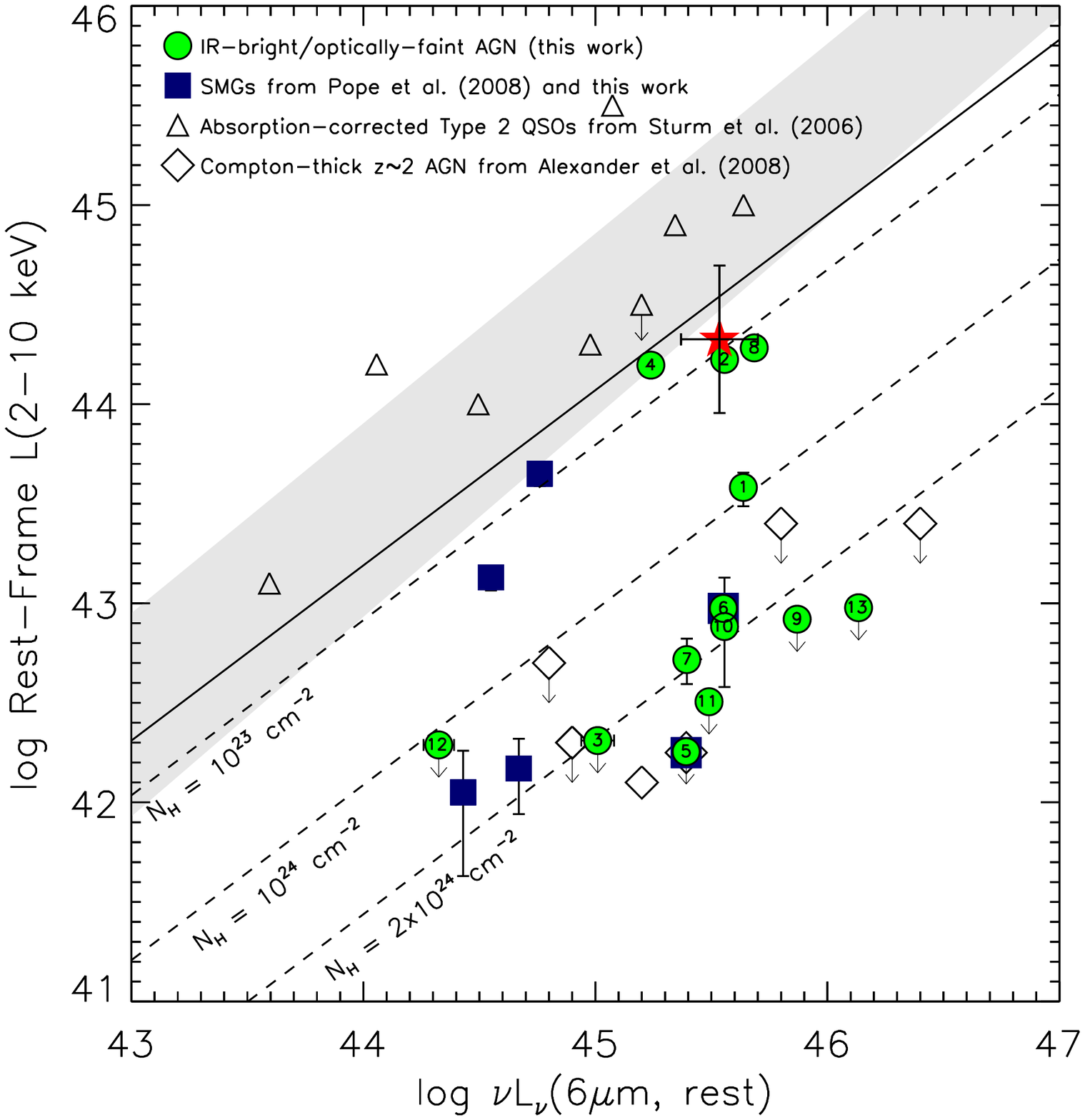}
\caption{Rest-frame obscured X-ray luminosity vs. rest-frame 6 \micron\ monochromatic luminosity, the values of which are given in Tables 2 and 4.   The \dogs\ are shown as circles.  The intrinsic relation for local AGN \citep{lutz04} and the intrinsic, luminosity-dependent relation of \cite{maiolino07}, which is based on both low and high redshift ($z\sim 2$) AGN, are given by the shaded region and solid line, respectively.   As indicated by the red star, the \textit{absorption-corrected} (e.g. intrinsic) luminosity ratio of the six sources in our sample with crude X-ray--derived column density estimates is broadly consistent with the \cite{maiolino07} and \cite{lutz04} relations (the error bars represent the standard deviation in the distribution of X-ray and 6 \micron\ luminosities).  Dashed lines give the expected observed relations for AGN obscured by column densities of $N_{\rm H} = 10^{23}, 10^{24}$, and $2 \times 10^{24}$~cm$^{-2}$, calculated assuming an intrinsic X-ray photon index of $\Gamma=2$.   For comparison,  the observed X-ray and intrinsic 6 \micron\ luminosities of the Compton-thick AGN from \cite{alexander08ct} are shown as diamonds and a sample of SMGs containing AGN drawn both from this work and from \cite{pope08smg} are shown as squares.  The absorption-corrected values for the Type II AGN from \cite{sturm06} are shown as triangles.  Based on this luminosity ratio, 5/13 \dogs\ are strong Compton-thick candidates with predicted columns of $N_{\rm H} \ge 2 \times 10^{24}$~cm$^{-2}$, and 5 additional sources lie near the Compton-thick/heavily obscured boundary ($N_{\rm H} \gsim 10^{24}$~cm$^{-2}$).}
\end{figure}

For comparison, we also show the absorption-corrected colors of the Type 2 QSOs from \cite{sturm06}, all of which fall in or above the \cite{lutz04} region, and the observed colors of the $z\sim2$ Compton-thick AGN from \cite{alexander08ct}, all of which fall below the $N_{\rm H} = 10^{24}$~cm$^{-2}$ relation\footnote{The source SMM J123600+621047 from \cite{alexander08ct} corresponds to source IRBG5 in this work. Because of the slightly different procedures used to estimate the rest-frame X-ray and IR luminosities, however, our calculated values of log~L$_{\rm {2-10 keV}} $(ergs~s$^{-1}$)$= 45.4$~ergs~s$^{-1}$ and log~$\nu L_\nu $(ergs~s$^{-1}$)$ < 42.3$ differ slightly from those in \cite{alexander08ct}: log~L$_{\rm {2-10 keV}} $(ergs~s$^{-1}$)$ = 45.3$ and log~$\nu L_\nu $(ergs~s$^{-1}$)$ < 42.4$.  For consistency, we plot the \cite{alexander08ct} source using the values measured here.}.  Furthermore, we also plot the observed colors of the six SMGs with a well-measured MIR AGN contribution from either \citet[C1, C3, GN04, GN06, GN07]{pope08smg} or this work (both IRBG5, which corresponds to source C1 in \cite{pope08smg}, and IRBG6 are SMGs).  For the \cite{pope08smg} sources, we estimate the intrinsic 6\micron\ luminosity using their Fig. 3 for which they performed spectral fitting similar to that described in \S4.  X-ray flux measurements were taken from \cite{alexander03} when available (C1 and GN04), and were calculated as described in \S5.1 otherwise.  Of the 6 SMGs, three (IRBG6, C3, GN04) are detected in the X-ray to high significance, two (GN06 and GN07) are weakly-detected at 6.2$\sigma$ and 3.3$\sigma$ above the background, respectively, and one (C1/IRBG5) remains X-ray undetected.

As shown in Figure 4, three of our 13 sources (IRBG2, IRBG4, and IRBG8) appear to be obscured yet Compton-thin ($N_{\rm H} \sim 10^{23}$~cm$^{-2}$); the remaining 10 lie near or above the Compton-thick boundary ($N_{\rm H} \gsim 10^{24}$~cm$^{-2}$).  This high fraction of Compton-thick candidates ($\sim 80\%$) is significantly larger than that found by \citet[][18\%]{lanzuisi09} thanks to the far deeper X-ray data ($T_{\rm X} =$200~ks to 2~Ms compared to $T_{\rm X} =5-70$~ks).   Furthermore, there is at least rough agreement between the column densities estimated above and those estimated here.  For instance, of the 3 sources with X-ray/MIR luminosity ratios indicative of column densities of log~$N_{\rm H}$~(cm$^{-2}$)~$ \lsim 23$, all have X-ray--derived column densities of $22 \le$~log~$N_{\rm H}$~(cm$^{-2}$)~$ \le 23$.  Of the remaining three sources with X-ray--derived column densities (IRBG1, IRBG6, and IRBG7), two are identified as Compton-thick candidates in Figure 4 and one (IRBG1) has an estimated column that falls just short of log~$N_{\rm H}$~(cm$^{-2}$)~$ = 24$.   While their X-ray--derived columns (log~$N_{\rm H}$~(cm$^{-2}$)~$ = 23.4-23.6$) place them just outside the Compton-thick regime, column densities based on low S/N flux ratios are often unable to distinguish Compton-thick AGN from their most heavily-obscured, yet Compton-thin, counterparts \cite[see, for example, Figure 3 of][]{alexander08ct}. On the basis of the X-ray and X-ray/MIR--derived column densities, we therefore conclude that all of the sources in our sample are likely to be X-ray obscured (log~$N_{\rm H}$~(cm$^{-2}$)$ \ge {22}$), and as many as 10 ($\sim 80\%$) may be Compton-thick.  Furthermore, at least 5 of these sources ($\sim 40\%$) lie at predicted column densities of $>2\times 10^{24}$~cm$^{-2}$, and are therefore strong Compton-thick candidates.  

One of these strong Compton-thick candidates, IRBG5, was previously identified as a potential Compton-thick quasar by both \cite{alexander08ct} and \citet{georgantopoulos09} and is also one of the SMGs plotted in Figure 4.  Its heavily obscured nature, however, is not unique among the sample of SMGs with measurable AGN contributions in the MIR.  In fact, 4 of the 6 SMGs plotted in Figure 4 ($\sim 70\%$) are Compton-thick candidates. Our results are therefore in agreement with those of \cite{alexander05}, who found that 80\% of SMGs containing X-ray--detected AGN are likely to be heavily obscured (log~$N_{\rm H}$~(cm$^{-2}$)~$ \gsim 23$) and potentially Compton-thick.  This similarity between the column density distributions of the \dog\ and SMG samples underscores the potential relationship between these two classes of objects, discussed in more detail in \S7 and \S8.

\subsection{X-ray Luminosity}

The observed and absorption-corrected X-ray luminosities of the \dogs\ are given in Table 4.  The mean, non-absorption-corrected 2-10 keV rest-frame luminosity of the 7 sources with soft-band detections (calculated as described in \S5.2) is log~$L_{\rm 2-10}$(ergs~s$^{-1}$)$=43.9$, which rises to log~$L_{\rm 2-10}$(ergs~s$^{-1}$)$=44.3$  when we apply the absorption corrections to the 6 sources with X-ray--derived $N_{\rm H}$ measurements.  

As discussed in \S5.2, an AGN's intrinsic X-ray luminosity can also be estimated from its rest-frame 6 \micron\ luminosity (see Figure 4).  This technique has the added advantage that it can be applied to all of the sources in the sample, regardless of whether they are formally detected in the X-ray.  Assuming a redshift of $z=2$ for the 3 sources lacking redshift estimates, we find a mean intrinsic 2-10 keV luminosity of log~$L_{\rm 2-10}$(ergs~s$^{-1}$)$=44.6 \pm 0.4$ for the 13 sources in our sample.  This value is only slightly higher than the estimate made above for the sources with X-ray--derived column density estimates (log~$L_{\rm 2-10}$(ergs~s$^{-1}$)$=44.3$), as would be expected if the flux ratio method underestimates the column densities of the most heavily obscured sources.  Furthermore, Figure 4 suggests that essentially all of the sources in our sample (with the exception of IRBG12, whose redshift is only $z=1.04$) have intrinsic X-ray luminosities that lie above the canonical division between low-luminosity Seyfert galaxies and high-luminosity QSOs, log~$L_{\rm x}$(ergs~s$^{-1}$)$= 44$.  The \dogs\ in our sample are therefore Type 2 QSOs, at least from the X-ray perspective.

\subsection{Correlation between X-ray and IR Properties}

All three of the sources in our sample that are optically-thick at 9.7 \micron\ ($\tau_{\rm 9.7 \micron} \sim 1-5$) are Compton-thick candidates with MIR/X-ray--estimated column densities of $N_{\rm H}\gsim2\times 10^{24}$~cm$^{-2}$.  The seven additional Compton-thick candidates, however, display little to no silicate absorption ($\tau_{\rm 9.7 \micron} \le 0.5$).   While it is possible that the 3 sources lacking solid redshift estimates (1 of which is a Compton-thick candidate) could have a silicate feature redshifted out of the observable band  ($z\ge2.6$), the highest confirmed redshift in our sample is that of IRBG13, $z=2.6$, a source with a prominent absorption feature.  Given the average sample redshift of $z=2$ and the low frequency of silicate absorption among the sources in our sample with confirmed redshifts, it therefore seems unlikely that the three sources lacking redshift information (and observable silicate features) lie at $z>2.6$ \textit{and} have significant silicate absorption.

It therefore appears clear that while silicate absorption is more likely amongst heavily X-ray--obscured \dogs, strong silicate absorption need not accompany X-ray obscuration.   A similar conclusion was drawn by \cite{brand08} for a sample of  luminous ($f_{\rm 24} > 2$~mJy) \dogs\ in the NOAO Deep Wide-Field Survey, as well as \cite{sturm05}, who show that bright ($F_{\rm 15 \micron} > 0.3$~mJy) X-ray--selected Type 2 AGN can show unabsorbed power-law continua in the IR despite heavy obscuration in the X-ray.  This apparent discrepancy between X-ray and MIR measures of obscuration could be due to an additional component of absorbing material that blocks the X-ray, but not the IR, emission region \citep[e.g.,][]{shi06}.  Alternatively, it could result from the filling-in of the silicate feature by an additional component of extended hot dust that lies beyond the obscuring torus, arising, for example, from the ionization cones as in NGC~1068, the prototypical Compton-thick AGN in the local Universe \citep{mason06,efstathiou95}.

\section{Star-formation Rates}

For the 7 \dogs\ with detectable PAH emission (i.e. AGN fractions $< 90\%$), we place limits on the star-formation rates (SFRs) using (1) the star-forming contribution to the observed-frame 24 \micron\ emission (in lieu of direct estimates based on the weakly-detected PAH emission features) and (2) the observed radio emission.

\subsection{24 \micron\ emission}

To measure the 24 \micron\ flux density arising from star-formation alone, we convolve the star-formation component of our IRS fit with the MIPS 24 \micron\ passband. We find that in the sources with detectable PAH emission, $\sim 25\%$ of the total 24 \micron\ flux density can be attributed to star-formation. We then use the redshift-dependent scaling relation of \cite{rieke09} to convert the observed MIR flux to a SFR, assuming the median sample redshift of $z=2$ for those sources lacking secure redshift estimates \citep[for further details, see][]{rieke09}.  With the exceptions of IRBG13, whose anomalously high SFR estimate of $2\times10^4~\Msun$~yr$^{-1}$ will be discussed below, and IRBG12, whose estimated SFR is $\sim 150 \Msun$~yr$^{-1}$, the resulting SFRs range from $\sim 1000-2000~\Msun$~yr$^{-1}$.  No meaningful constraints can be placed on the SFRs of those sources with AGN contributions in excess of 90\% (SFR$\le 7000-5\times 10^4~\Msun$~yr$^{-1}$).  

These estimates, however, are based on a relationship derived from local starburst, LIRG, and ULIRG templates.  At high-redshift, luminous star-forming galaxies exhibit strong PAH features typical of less-luminous local galaxies, due perhaps to extended star-formation or lower metallicity \citep[e.g.,][]{sajina07,papovich07,pope08smg,rigby08,farrah08,murphy09,menendez09}.  Local templates are therefore likely to overestimate the intrinsic 24 \micron\ flux (and therefore the SFR) of high-z ULIRGS significantly at $L(TIR) = 10^{13}~\Lsun$, the typical luminosities of luminous \dogs\ \citep[see][]{rieke09,tyler09}.  Additional uncertainties in the derived SFRs arise from the lack of luminous $L({\rm TIR})> 2 \times 10^{12}~\Lsun$ ULIRGS in the local Universe from which to calibrate the SFR relation.  The IR-derived SFRs of the \dogs\ should therefore be taken as rough upper limits on the true SFRs.

\subsection{Radio emission}
 
To place an independent constraint on the SFRs, we therefore consider the 1.4~GHz radio flux density.  Of the 7 \dogs\ with detectable PAH emission, 6 have radio counterparts in the catalogs of \cite{richards00}, \cite{ivison07}, or \cite{kellermann08}. Assuming a radio spectral index of $\alpha = -0.7$, where $f_{\rm \nu} \propto \nu^{\alpha}$, we K-correct the observed radio flux and use the relation of \cite{rieke09} to estimate the SFR.  With the exception of IRBG13, whose radio-derived SFR falls just short of the IR-derived SFR, the radio SFRs exceed the IR SFRs by factors of $\sim 1.2-10$, suggesting that a significant fraction of the radio flux density comes not from star-formation, but from AGN activity.  Therefore, the radio-derived SFRs again represent only an upper limit on the true star-formation activity.

To isolate the star-forming contribution to the radio emission, we therefore turn to the radio-infrared correlation, characterized by the parameter $q_{\rm 24} =$ log$(f_{\rm 24 \micron}/f_{\rm 1.4 GHz})$.   By using this correlation together with the measured star-forming contribution to the 24 \micron\ flux density to calculate the star-forming radio flux density and then the SFR, we again utilize the MIR emission as a proxy for the SFR.  This method, however, allows us to calibrate the SFR using the observed radio/SFR correlation in place of the 24\micron/SFR correlation discussed above. 

A number of different studies have attempted to quantify $q_{\rm 24}$ and have returned values ranging from 0.52 \citep{beswick08} to 1.39 \citep{boyle07}.  Here, we adopt the luminosity-dependent $q_{\rm 24}$ relation of \cite{rieke09} which is based on the local IRAS bright galaxy sample (BGS) and for which $q_{\rm 24} = (-1.275 \pm 0.756)+(0.224\pm0.066) \times $~log~$L(TIR)/\Lsun$ at log~$L(TIR)/\Lsun > 11$.  At log~$L(TIR)/\Lsun = 13.0$, the typical total infrared luminosity of luminous \dogs\ \citep{tyler09}, $q_{\rm 24} = 1.64$.

The observed value of $q_{\rm 24}$ also varies with redshift as the IR spectral features pass through the 24 \micron\ band.  To correct for this variation, we adopt the observed $q_{\rm 24}$ redshift evolution of \cite{ibar08}, who found that $q_{\rm 24} = (0.85 \pm 0.01) + (-0.20 \pm 0.01) \times z$.  Not surprisingly, this observed redshift evolution is reproduced by the log~$L(TIR)/\Lsun \sim 12.0$ templates of \cite{rieke09}.  After scaling the \cite{ibar08} relationship upwards to match the \cite{rieke09} result at $z=0$, we used the resulting correlation and our estimate of the star-forming 24 \micron\ flux density to determine the expected radio emission from star-formation.

We find that star-formation accounts for 3\% - 23\% of the total radio emission in our sources.  Again with the exceptions of IRBG12 and IRBG13, whose estimated SFRs are $\sim 1600~\Msun~yr^{-1}$ and $20~\Msun~yr^{-1}$, respectively, we derive SFRs of $\sim 200-400~\Msun~yr^{-1}$ using the star-formation-dominated radio emission.  Because the high-z ULIRGS/HyperLIRGS in our sample are likely to have star-forming emission similar to that of local LIRGS/ULIRGS (see \S6.1), however, the true intrinsic value of $q_{\rm 24}$ is likely to be lower than the value assumed above. The estimated SFRs can be therefore be taken as lower-limits, subject of course to the intrinsic scatter in the radio/IR correlation.

\subsection{The high SFRs of luminous IR-bright/optically-faint galaxies}  

To summarize, 5 of the 7 \dogs\ in our sample with detectable PAH emission and/or radio counterparts have estimated SFRs in the range of $\sim 200$ to $2000 \Msun~yr^{-1}$.  The SFR of IRBG12 ($\sim 20 - 150~\Msun~yr^{-1}$) is lower than the others because of its significantly lower redshift ($z=1.04$).  The SFR of IRBG13, however,  is anomalously high: $1.6-2.0\times 10^4~\Msun~yr^{-1}$.  This overestimation likely stems from the fact that both an AGN+SF fit \textit{and} an AGN-only fit reasonably reproduce the IRS spectrum ($\chi^{2}_\nu$ = 0.79 and 0.85, respectively).   While potential AGN-only fits likewise exist for many of the sources in our sample, IRBG13 has the smallest offset in the reduced $\chi^{2}$ values of the AGN+SF and AGN-only fits and is therefore the source for which the origin of the IR emission is most uncertain. 

The SFRs derived above are clearly approximate in nature. Nevertheless, two of the 7 sources examined above (IRBG5 and IRBG6) are known SMGs, heavily dust-enshrouded sources with typical SFRs of $500-2000 \Msun~yr^{-1}$ \citep{alexander05,swinbank04}.  As their IR properties are typical of the rest of our sample, it therefore seems likely that our estimated SFRs of $200-2000 \Msun~yr^{-1}$ reasonably approximate the high level of star formation present in luminous \dogs. Furthermore, our estimated SFRs are in agreement with the results of \cite{tyler09}, who find via far-infrared measurements that approximately a third of \dogs\ similar to those selected here have indications of star formation at a level that produces a substantial fraction of their bolometric luminosities, which are typically 10$^{13}$ L$_\odot$ \citep{tyler09}.  Finally, such SFRs are also in agreement with the predictions of \cite{desika09dog}, who find that luminous \dogs\ are best modeled by gas rich mergers undergoing rapid AGN growth and/or star-formation at rates of $\sim 500-1000 \Msun$~yr$^{-1}$.  

Presumably, the objects with detected star formation represent the high end of a continuous luminosity function, where the lower levels of star formation are masked by the AGN outputs. That is, it is likely that these sources are in general characterized by star formation in the high-LIRG to ULIRG range, even though their MIR (14-37 \micron) luminosities are AGN-dominated. 

\begin{figure*}
\epsscale{0.8}
\plotone{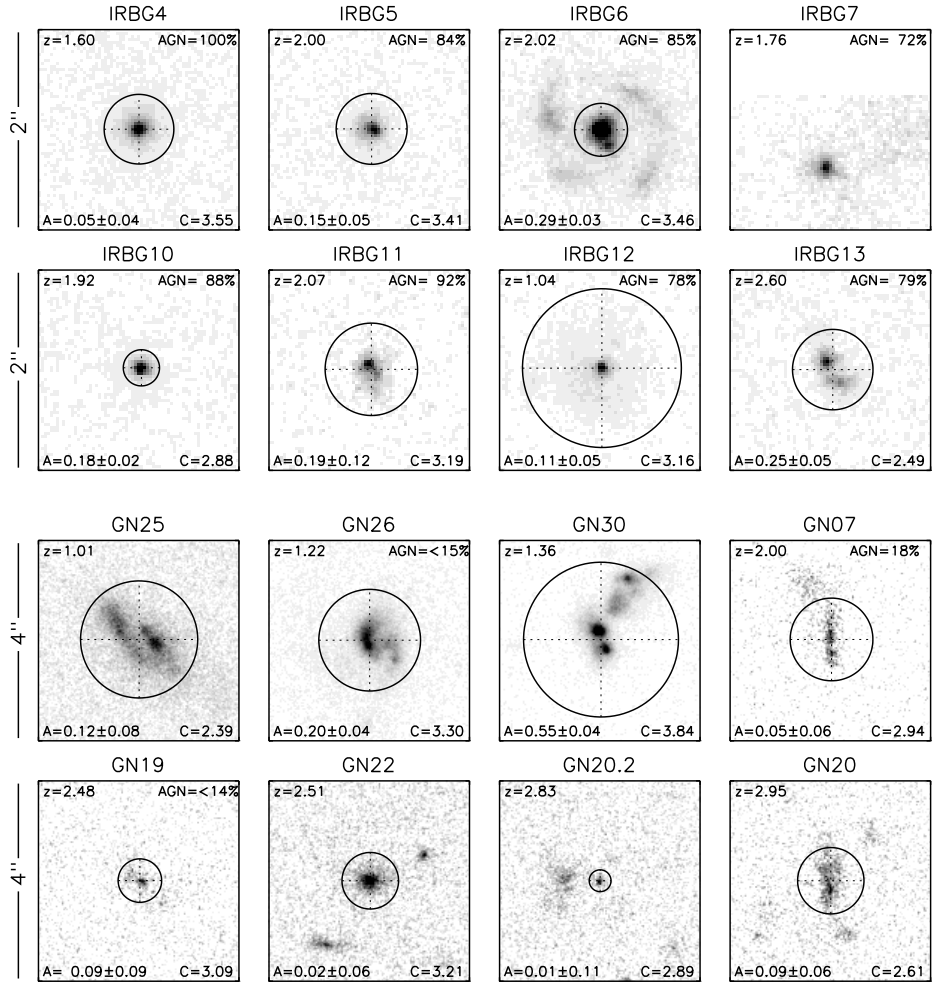}
\caption{GOODS and AEGIS ACS $i$-band images of the 8 \dogs\ with deep ACS coverage (top, image diameter of 2\arcsec) and the 8 GOODS-N SMGs from the literature with similar redshifts and $i$-band magnitudes (bottom, image diameter of 4\arcsec).  The redshift and IRS-derived AGN fraction are given in the top left and right corners of the image, and the asymmetry and concentration indices for each galaxy are given in the bottom left and right corners of each image, respectively.  The redshifts and AGN fractions for the SMGs come from \cite{pope06,pope08smg} and \cite{murphy09}.  The circle gives the measured Petrosian radius, and the cross-hairs indicate the rotation center chosen by the asymmetry algorithm. Because it extends beyond the edge of the image, we are unable to estimate a reliable Petrosian radius or morphological indices for IRBG7.}
\end{figure*}

\section{Host Galaxy Morphologies}

Because of the $\sim 30\%$ overlap of IRBG and SMG samples, as well as their similar redshift distributions, surface densities, and clustering properties, several authors have suggested that luminous \dogs\ may evolve from SMGs \citep[e.g.,][]{dey08,pope08dog,brodwin08,bussmann09,desika09smg,desika09dog}.  Furthermore, early studies of SMGs suggested that $61\% \pm 21\%$ may be in the early stages of a major merger \citep{conselice03smg}, although, on the basis of asymmetry measurements alone, more recent work has suggested that SMGs are no more likely to be undergoing major mergers than lower-luminosity high-redshift galaxies \citep{swinbank10}.  While this work has called into question the uniqueness of the role that mergers play in these systems, the merger-driven origin of both SMGs and luminous \dogs\ is supported by the {\sc gadget2} simulations of \cite{desika09smg,desika09dog}, who find that both samples naturally arise from gas-rich mergers, with SMGs representing the early, star-formation dominated phase of the merger, and \dogs\ representing the later phase of final coalescence in which star-formation and obscured black hole growth both proceed at rapid rates (see Figure 1 in \citealt{hopkins08}).  If this scenario is correct, luminous \dogs\ should therefore exhibit signs of recent merger activity, although one might expect them to be more dynamically relaxed than their SMG predecessors.

\subsection{Comparison between \dogs\ and SMGs}

Previous morphological studies of \dogs\ have used targeted \textit{HST} and Keck AO observations in the Bootes Field of the NOAO Deep Wide-Field Survey to (1) measure their typical sizes and shapes via Sersic fitting and {\sc GALFIT} galaxy decomposition and to (2) characterize their morphologies on the basis of the Gini, $M_{\rm 20}$, and concentration indices \citep{bussmann09, melbourne09}.  They conclude that \dogs\ are preferentially found in disk-like galaxies, and only $\sim 15\%$ show multiple resolved components indicative of ongoing mergers, far lower than the $\sim 50\%$ interaction rate found for the average $z=2$ ULIRG and the $\sim 60\%$ merger rate estimated for SMGs \citep{conselice03smg,dasyra08,bussmann09,melbourne09}.  However, \dogs\ do appear to have smaller projected sizes ($r_{\rm Petrosian} = 0.5\arcsec - 1.5\arcsec$) than SMGs ($r_{\rm Petrosian} = 0.5\arcsec - 2.5\arcsec$), suggesting that they may in fact represent a later, more relaxed phase of a major merger \citep{bussmann09}.

To expand upon these studies and further test whether \dogs\ are the product of major-mergers, we directly compare below the morphologies of \dogs\ and SMGs in a self-consistent manner.  Because IRBG5 and IRBG6 are confirmed SMGs, we treat these two sources separately in the analysis below.  We also note that because the EGS does not yet contain deep sub-mm coverage, it is possible that one or more of the EGS \dogs\ could also be SMGs.  \cite{pope08dog}, however, found an SMG fraction of only 30\% amongst luminous \dogs.  As the fraction of SMGs in our sample of 8 \dogs\ with deep ACS coverage is already $\sim 25\%$,  we therefore expect little to no additional contamination by SMGs in our morphological sample.

We select as a comparison sample of SMGs those sources in the GOODS-N field \citep{pope05} with redshifts and apparent magnitudes consistent with those of the \dogs: $1 < z < 3$ and $I<26$, and confirm that none are also \dogs\ (although GN25 has a high $f_{\rm 24}/f_{\rm R}$ of $\sim 900$; \citealt{cowie04,pope06}).  The observed-frame $i$-band images of the 8 \dogs\ and 8 SMGs with deep ACS imaging from GOODS or the EGS are shown in Figure 5.

\subsubsection{Merger Fraction}

To quantitatively compare the morphologies of these two samples, we calculate the Petrosian radii and concentration $(C)$ and asymmetry $(A)$ indices using the approach of \citet{conselice00} and the error-estimation method of \citet{shi09}, and confirm excellent agreement between our measured parameters and those of \citet{shi09} for the two sources that lie in both samples:  IRBG5 and IRBG6 (Y. Shi 2009, private communication).

Of these parameters, the asymmetry index is the most relevant as it tends to be high ($A>0.35$ in the local Universe) when merging galaxies are undergoing either their first pass or their final coalescence \citep{conselice03mergers,lotz08}.  We caution, however, that this index is designed to measure the asymmetry within the Petrosian radius ($r_{\rm Petrosian}$) of the primary source.  If the merging galaxies are separated by a large distance (as occurs, for example, during the merger stages between first pass and final coalescence), the secondary source will fall outside of the Petrosian radius of the primary source and the index will be low.  Furthermore, surface brightness dimming causes the observed asymmetry to decrease with increasing redshift.   While the correction for this effect is somewhat uncertain, \cite{conselice05} estimate an offset of $\Delta A = -0.15$ between $z=0$ and $z=2$ for irregular galaxies.  As the \cite{shi09} error method results in measured asymmetries 0.05 lower than those found by the minimum error method of \citet{conselice00}, we therefore adopt a merger index of $A \ge 0.15$ for our $z\sim2$ sample\footnote{We note that \cite{shi09} find a slope ($\Delta A/\Delta z$) between $z=0$ and $z=1$ that is twice that of \cite{conselice05} when using local LIRGs as input.  It therefore remains possible that galaxies in our sample with $A < 0.15$ are undergoing mergers as well.}.  

The measured concentration and asymmetry indices are shown in Figure 6.  By these measures alone, the \dogs\ and SMGs appear to have similar morphologies.  Furthermore, all of the sources with clear double nuclei have $A > 0.15$, as expected.  Several SMGs, however, do indeed have nearby counterparts that lie outside the Petrosian radius of the primary source.  If we therefore count as mergers all sources with either $A > 0.15$ or nearby counterparts that fall outside $r_{\rm Petrosian}$ (e.g. GN22,GN20.2,GN20), we find that 5 of the 7 ($\sim 70\%$) \dogs\ and 5 of the 8 ($\sim 60\%$) SMGs are merger candidates.

\begin{figure}
\epsscale{1.06}
\plotone{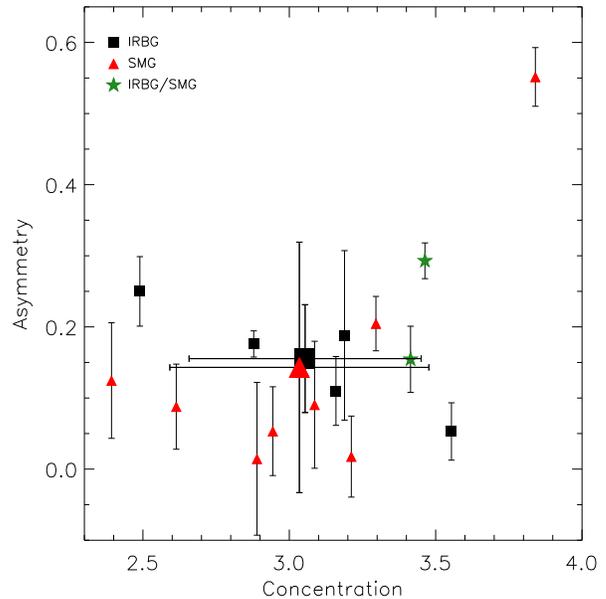}
\caption{Asymmetry vs. concentration indices for the \dogs\ (black squares), SMGs (red triangles), and sources that meet both criteria (green stars).  The mean properties and standard deviations of the \dog\ and SMG samples are shown by large symbols.  As noted in \S7.1.1, the measured asymmetry will be low when merging galaxies lie at large separations, as is the case for many of the SMGs.  To calculate a merger fraction, we therefore consider both the measured asymmetry and the fraction of sources with nearby neighbors.  }
\end{figure}

It is also worth noting that two of the three \dogs\ which are optically thick at 9.7 \micron\ (IRBG11 and IRBG13) are strong merger candidates with large asymmetries and clearly disturbed morphologies.  The third, IRBG7, falls too near the edge of the ACS image to accurately calculate the asymmetry.  Nonetheless, it appears to lie at the edge of a large patch of diffuse emission, potentially due to an ongoing merger (see Figure 5).  As discussed in \S5.4, these three optically-thick sources are the only Compton-thick candidates with strong silicate absorption.  It therefore appears plausible that ongoing mergers are at least partially responsible for the high degree of MIR obscuration in these heavily X-ray--obscured \dogs.  In their sample of local ULIRGs, \citet{veilleux09} similarly find that the depth of the 9.7 \micron\ silicate feature is generally largest during the close pre-merger and merger stages.  They caution, however,  that this correlation suffers from a large degree of scatter, likely due to variations in both the initial conditions (e.g., the structure, surface mass density, and gas fractions of the infalling galaxies) and specific characteristics (e.g., orbital geometry and mass ratio) of the mergers themselves.  This scatter may explain why the three remaining merger candidates (IRBG5, IRBG6, and IRBG10) show little to no MIR obscuration ($\tau_{\rm 9.7 \micron} = 0.0-0.3$).

\subsubsection{Physical Size}

While a majority of both SMGs \textit{and} \dogs\ appear to be sufficiently disturbed to be classified as major mergers, these two samples do appear to differ in their physical size, as inferred by \cite{bussmann09}.  A histogram of the Petrosian radii in physical units of kpc is shown in Figure 7.  The mean $r_{\rm Petrosian}$ of the SMGs, $6.8 \pm 3.7$~kpc, is nearly twice that of the \dogs, $3.7 \pm 1.8$~kpc.  If we exclude the 3 lowest-redshift SMGs with $z<1.4$, the mean $r_{\rm Petrosian}$ of the SMGs drops to $4.5 \pm 2.0$~kpc, suggesting that the apparent offset in physical size may be driven at least in part by evolution in the SMG population.  \citet{swinbank10}, however, detect no significant size evolution in either the optical or NIR radii of a sample of 25 SMGs.  Furthermore, they find a mean optical (NIR) Petrosian radius of $6.9 \pm 0.7$~kpc ($7.7 \pm 0.6$~kpc) for their SMG sample, consistent with our results and again a factor of $\sim$2 higher than observed for the \dogs.  

\begin{figure}
\epsscale{1.0}
\plotone{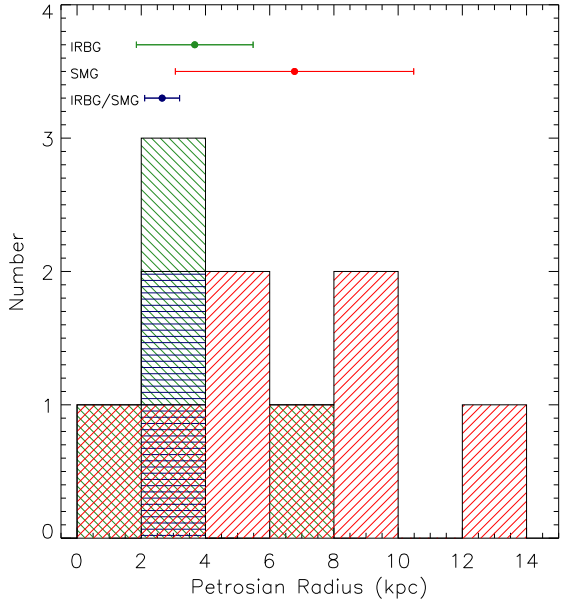}
\caption{Petrosian radii of the \dogs\ (green histogram, angle=$-45^{\circ}$), SMGs (red histogram, angle=$45^{\circ}$), and sources that meet both criteria (blue histogram, angle=$0^{\circ}$).  The mean radii (and standard deviations) for the three samples are shown at the top of the plot.}
\end{figure}

This apparent offset in size may be a natural byproduct of the merger hypothesis, as the Petrosian radius of a merger remnant peaks strongly during both the first pass and the final merger stages and can easily obtain a value twice that of the pre-merger, maximal separation (when only one galaxy is contained within the Petrosian radius), and post-merger stages \citep[][J. Lotz 2009, private communication]{lotz08}.  An offset in $r_{\rm Petrosian}$ of a factor of $\lsim 2$ is therefore consistent with the hypothesis that SMGs are observed in the first stages of a major merger whereas \dogs\ represent a post-merger phase following the final coalescence.  What remains more difficult to explain, however, are the simultaneously \textit{large} asymmetries and \textit{small} radii of the \dogs, as these two parameters should both peak at approximately the same stages during the merger.  Larger samples of such sources are clearly needed to resolve this apparent discrepancy.

\section{Discussion}

Several authors have suggested that \dogs\ and SMGs may be evolutionarily related, each representing a different phase in the major-merger scenario originally proposed by \citet{sanders88} \citep[e.g.][]{dey08,pope08dog,coppin10,desika09smg,desika09dog,brodwin08}.  To test this hypothesis, we explore below the agreement between the properties of the \dog\ and SMG samples, as measured here and in the literature, and the major-merger model.

The major-merger (e.g., 3:1 or greater mass ratio) scenario, as presented by \citet{hopkins08}, proceeds as follows. During the merging galaxys' first pass, star-formation activity begins to increase with concurrent AGN activity occurring only under certain orbital geometries and/or when the host galaxys' discs are particularly unstable.  As the galaxies undergo final coalescence, however, massive inflows power both strong starburst activity and heavily-obscured, yet rapid, black hole growth.  Eventually, feedback from the black hole and from supernovae disperse the remaining gas in a brief ``blowout'' phase characterized by a dust-obscured, yet Type 1, AGN. Finally, when the gas and dust have been fully dispersed, the source appears as an unobscured quasar (unless viewed through the canonical torus). The general picture is therefore one in which a major-merger fuels both star-formation and black hole growth (thus leading to the observed $M_{\rm BH}/\sigma$ relation), with star-formation dominating at early times, and AGN activity at later times. 

Before continuing, it is worth noting that this model is almost certainly an oversimplification.  As emphasized by \citet{veilleux09}, the internal structure of the merging galaxies, the orbital parameters (e.g., prograde or retrograde), the mass ratio, and sub-resolution physics will ultimately determine the evolutionary path followed by any one merging galaxy pair, leading to variations on the model presented above and scatter in the observed correlations.  Nonetheless, the general picture presented above should hold \textit{on average} if this process is responsible for the SMG and \dog\ phenomena. 

In the context of this model, SMGs are assumed to represent the early, star-formation dominated phase of the major merger, and \dogs\ the later, AGN-dominated phase.  To first order, this progression is supported by the fact that while $\sim 85\%$ of SMGs are star-formation dominated in the MIR \citep{pope08smg,hainline09,coppin10}, and nearly all luminous \dogs\ are AGN-dominated in the MIR, there is an $\sim 30\%$ overlap in the SMG and \dog\ populations, at least for samples extending down to $f_{\rm 24} \ge 100$ \microjy\ \citep{pope08dog}.  Furthermore, the SMGs that also meet the \dog\ criterion tend have the largest AGN contributions, and are therefore likely candidates to be undergoing a transition from star-formation to AGN dominance \citep[e.g.,][]{pope08dog,coppin10}.

In addition, while SMGs tend to be star-formation-dominated in the MIR, $\sim$28-50\% host obscured AGN \citep[][also see Figure 4]{alexander05} as predicted for the subset of mergers with initial conditions and orbits conducive to early (obscured) black hole growth.  Furthermore, at least $\sim$30-40\% of luminous \dogs\ have ULIRG-level SFRs \citep[][this work]{tyler09}.  As rapid star-formation is predicted throughout the obscured AGN phase of a major merger, this too is consistent with the major merger scenario. 

This general picture is also supported by the work of \citet{alexander08smg}, \citet{dey08}, and \citet{brodwin08}. For instance, \citet{alexander08smg} find that black hole growth in SMGs lags behind the growth of the host galaxy, contrary to what is seen in high-redshift optical (e.g., unobscured) and radio-selected samples of quasars  \citep[e.g.,][]{peng06,mclure06}.  This finding suggests that SMGs are at an earlier evolutionary stage than their unobscured, AGN-dominated counterparts, again in agreement with the merger scenario.  More generally, \citet{dey08} find that the surface densities and redshift distributions of $f_{\rm 24} \ge 300$~\microjy\ \dogs\ \citep[at least $\sim 60\%$ of which are likely to be AGN-dominated in the MIR,][]{pope08dog} are comparable to those of luminous SMGs \citep[$F_{\rm 850 \micron} > 6$~mJy,][]{chapman05,coppin06}. Similarly, \cite{brodwin08} analyze the clustering properties of \dogs\ and find that $f_{\rm 24} \ge 300$~\microjy\ \dogs\ have a correlation length comparable to that of SMGs \citep{blain04}.  The most luminous ($f_{\rm 24} > 700$~\microjy) \dogs, however, may be more strongly clustered, and thus lie in richer environments, than both their lower MIR-luminosity counterparts and the typical SMG. 

While no one of these arguments confirms the potential merger origin for SMGs or \dogs, collectively, the results suggest that the merger scenario first proposed by \citet{sanders88} may provide an acceptable explanation for the properties of these two samples.  Ideally, one might hope to either confirm or reject this hypothesis on the basis of the observed morphologies.  Unfortunately, however, the morphological properties of these samples remain amongst their most uncertain characteristics.  As discussed in \S7.1, the asymmetry measure, commonly used as a merger indicator in the local Universe, suffers at high redshift from surface brightness dimming, a $(1+z)^4$ effect, and is insensitive to intermediate merger stages in which the galaxies lie at large separations.  When using this measure alone, \citet{swinbank10} therefore conclude that SMGs are no more likely to appear as mergers than a bolometrically less luminous sample of high redshift star-forming galaxies.  Optical (e.g., $i$-band) measurements of this and other objective morphology indicators are also complicated by the fact that at $z\sim2$, the observed-frame optical emission probes the the rest-frame UV emission, which tends to be more strongly clumped.  However, when morphologies of AGN-dominated \dogs\ are measured at longer wavelengths, the contribution from the unresolved AGN emission, now less obscured, can likewise bias the results \citep{bussmann09,melbourne09}.

Prior studies of \dog\ morphology have therefore turned to the fraction of multiple resolved components as a measure of ongoing mergers, and have found a rate of only $\sim 15\%$ for \dogs\ \citep{bussmann09,melbourne09}, far lower than is seen in local and high redshift samples of ULIRGs.  However, this measure too is biased, and will not identify sources in the latest stages of a merger.  We therefore chose to combine the asymmetry \textit{and} near-neighbor/multiple-component methods, and conclude that the majority of both SMGs \textit{and} \dogs\ are likely to be undergoing mergers. Furthermore, the smaller physical size of the \dogs, a trend first noted by \citet{bussmann09} and confirmed in \S7.1.2, suggests that they are likely in a later, more relaxed merger stage, a hypothesis also supported by the AGN and star-forming characteristics presented above.   Clearly, however, one would like to confirm this finding using a far larger, deeper, and higher-resolution sample, one that will be provided by the upcoming CANDELS (Cosmic Assembly Near-IR Deep Extragalactic Legacy Survey) \textit{Hubble} WFC3 NIR imaging survey of the GOODS, COSMOS, EGS, and UDS fields.

\section{Summary}

In summary, we present the X-ray, star-forming, and morphological properties of a sample of luminous ($f_{\rm 24} \gsim 700$~\microjy) IR-bright/optically-faint galaxies (\dogs, $f_{\rm 24}/f_{\rm R} \gsim 1000$) selected in deep X-ray fields.  Our major findings are as follows:

\begin{itemize}

\item The infrared colors and IRS spectra confirm that all of the $f_{\rm 24} \gsim 700$\microjy\ \dogs\ are AGN-dominated. 

\item All of the sources appear to be X-ray obscured, and 4 remain X-ray non-detected despite deep X-ray coverage.  As many as $\sim 40\%$ are strong Compton-thick candidates, and $\sim 40\%$ more appear to lie near the Compton-thick/heavily obscured (N$_{\rm H}$ = 10$^{24}$ cm$^{-2}$) boundary.
 
\item The mean intrinsic X-ray luminosity of the sample is log~$L_{\rm 0.5-8}$(ergs~s$^{-1}$)$\sim44.6\pm0.4$, and all but one source has a MIR-derived intrinsic X-ray luminosity in excess of log~$L_{\rm 0.5-8}$(ergs~s$^{-1}$)$=44$, the canonical dividing line between Seyfert galaxies and QSOs.  These sources are therefore Type-2 QSOs from the X-ray perspective.

\item In general, the sources with the largest X-ray obscuration are those most likely to exhibit strong silicate absorption.  Furthermore, at least two of the three \dogs\ that are optically thick at 9.7 \micron\ are likely to be undergoing mergers.  However, silicate absorption need not accompany X-ray obscuration, and merger candidates are observed that lack strong silicate features. 

\item 30-40\% of \dogs\ have star formation rates in the ULIRG range; presumably the rest of the sample also has relatively high levels of star formation that are masked by the AGN outputs.
 
\item As many as $\sim 70\%$ of the \dogs\ in our sample show asymmetries indicative of mergers, similar to the incidence of mergers in SMGs at similar redshift. However, the IR-bright/optically faint sources tend to be more compact than the SMGs, suggesting that they may be in a later stage of merging.
 
\item These characteristics are consistent with the proposal that these objects represent a later evolutionary stage following soon after the star-formation-dominated one represented by the SMGs. 

\end{itemize}

\acknowledgments
JLD thanks STScI for support through the Giacconi Fellowship, and Caltech/JPL for support through contract 1255094 to the University of Arizona, and DMA thanks the Royal Society and Leverhulme Trust for a University Research Fellowship and Philip Leverhulme Prize, respectively. P. G. P.-G. acknowledges support from the Spanish Programa Nacional de Astronom\'{\i}a y Astrof\'{\i}sica under grants AYA 2006-02358 and AYA 2006-15698-C02-02, and from the Ram\'on y Cajal Program financed by the Spanish Government and the European Union.  Finally, we thank Alex Pope, Elise Laird, Yong Shi, Jen Lotz, Laura Hainline, and the anonymous referee for helpful discussions and suggestions that improved the paper.

\end{document}

%% file: table1.tex
\begin{centering}
\begin{deluxetable*}{llllccccc}
\centering
\tabletypesize{\footnotesize}
\tablewidth{0pt}
\tablecaption {Observations}
\tablehead{
\colhead{ID}&
\colhead{Field}&
\colhead{Program}&
\colhead{Observation}&
\multicolumn{2}{c}{Ramp Duration (s)}&
\multicolumn{2}{c}{Number of Cycles}&
\colhead{Total} \\
\colhead{}&
\colhead{}&
\colhead{ID \tablenotemark{a}}&
\colhead{Date}&
\colhead{LL1\tablenotemark{b}}&
\colhead{LL2\tablenotemark{c}}&
\colhead{LL1\tablenotemark{b}}&
\colhead{LL2\tablenotemark{c}}&
\colhead{Integration (s)\tablenotemark{d}}
}
\startdata
IRBG1     & E-CDFS   & 30419     & 2006 Sept 15    & 120  & 120  & 10  & 10  &  4800  \\
IRBG2     & CDF-S     & 30419     & 2007 Aug  31    & 120  & 120  & 23  & 23  & 11040  \\
IRBG3     & E-CDFS   & 30419     & 2006 Sept 14    & 120  & 120  & 27  & 27  & 12960  \\
IRBG4     & CDF-S    & 30419     & 2007 Aug  31    & 120  & 120  & 15  & 15  &  7200  \\
IRBG5     & CDF-N   & 20456\tablenotemark{e}     & 2006 Apr  24    & 120  & 120  &  6  & 6   &  2880  \\
IRBG6     & CDF-N  & 20733     & 2006 May  26    & 30   &  30  & 80  & 80  & 12880  \\   
IRBG7     & CDF-N  & 20456\tablenotemark{f}     & 2006 Apr  22    & 120  & 120  &  6  & 6   &  9600  \\
IRBG8     & EGS       & 30419     & 2007 Apr  27    & 120  & 120  & 12  & 12  &  5760  \\
IRBG9     & EGS       & 30419     & 2006 Jun  26    & 120  & 120  &  4  & 4   &  1920  \\
IRBG10   & EGS       & 30419     & 2007 Apr  27    & 120  & 120  &  7  & 7   &  3360  \\
IRBG11   & EGS       & 30419     & 2007 Mar  25    & 120  & 120  & 28  & 28  & 13440  \\
IRBG12   & EGS       & 30419     & 2007 Mar  25    & 120  & 120  & 18  & 18  &  8640  \\
IRBG13   & EGS       & 30419     & 2007 Apr  27    & 120  & 120  &  6  & 6   &  2880  \\
\enddata
\tablenotetext{a}{The PI's of programs 30419, 20456, and 20733 are G. H. Rieke, R. Chary, and C. M. Urry, respectively.}
\tablenotetext{b}{Long Low 1st order, 19.5-38.0 \micron}
\tablenotetext{c}{Long Low 2nd order, 14.0-21.3 \micron}
\tablenotetext{d}{The IRS nodding performed in Standard Staring mode results in two spectra per cycle, each with an exposure time equal to the ramp duration.}
\tablenotetext{e}{The IRS spectrum for this source was previously published in \cite{pope08smg} and \cite{murphy09}}
\tablenotetext{f}{The IRS spectrum for this source was previously published in \cite{murphy09}}
\end{deluxetable*}
\end{centering}

%% file: table2.tex
\begin{centering}
\begin{deluxetable*}{lrrllcccccccccc}
\centering
\tabletypesize{\footnotesize}
\tablewidth{0pt}
\tablecaption {IR-Bright/Optically-Faint Sample}
\tablehead{
\colhead{ID}&
\colhead{RA (J2000)}                     &
\colhead{DEC (J2000)}                    &
\colhead{$z_{\rm spec}$\tablenotemark{a}}            &
\colhead{$z_{\rm IRS}$\tablenotemark{b}}            &
\colhead{$f_{\rm 24}$}            &
\colhead{$R (AB)$}                &
\colhead{$f_{\rm 24}/f_{\rm R}$}   &
\colhead{$\alpha_{\rm IRAC}$\tablenotemark{c}}   &
\colhead{$\alpha_{\rm IRS}$\tablenotemark{c}}   &
\colhead{AGN\%\tablenotemark{d}}   &
\colhead{$f_{\rm 24,SF}$\tablenotemark{e}}   &
\colhead{$\tau_{\rm 9.7 \micron}$\tablenotemark{f}}   &
\colhead{$L_{6\micron}$\tablenotemark{g}}   \\
\colhead{} &
\colhead{} &
\colhead{} &
\colhead{} &
\colhead{} &
\colhead{$(\mu \rm{Jy})$} &
\colhead{} &
\colhead{} &
\colhead{} &
\colhead{} &
\colhead{} &
\colhead{$(\mu \rm{Jy})$} &
\colhead{} &
\colhead{} 
}
\startdata
  IRBG1  &03:31:46.6  &-27:45:53.0  &   \nodata                      &           \nodata  &    1069  &    25.3  &    3961  &     -1.92  &   -2.08  &                96             $\pm$ 2  &   41  &  0.0  & 45.6 \\
  IRBG2  &03:31:58.3  &-27:50:42.2  &   \nodata                      &    2.28$\pm$ 0.06  &     851  &    25.2  &    2934  &     -1.86  &   -2.43  &                94             $\pm$ 5  &   99  &  0.5  & 45.6 \\
  IRBG3  &03:32:10.5  &-28:01:09.2  &   \nodata                      &    1.62$\pm$ 0.02  &     751  &    25.1  &    2208  &     -1.46  &   -2.06  &                71             $\pm$ 4  &  286  &  0.4  & 45.0 \\
  IRBG4  &03:32:37.8  &-27:52:12.5  &      1.60                      &           \nodata  &    1066  &    24.8  &    2370  &     -1.78  &   -2.07  &               100             $\pm$ 3  &    0  &  0.0  & 45.2 \\
  IRBG5  &12:36:00.2  & 62:10:47.3  &      2.00                      &    2.00$\pm$ 0.03  &    1144  &    24.0  &    1204  &     -2.13  &   -2.06  &                84             $\pm$ 3  &  293  &  0.1  & 45.4 \\
  IRBG6  &12:36:35.6  & 62:14:23.9  &      2.02                      &    2.04$\pm$0.004  &    1497  &    23.8  &    1426  &     -2.00  &   -2.04  &                85             $\pm$ 1  &  351  &  0.0  & 45.6 \\
  IRBG7  &12:37:26.5  & 62:20:27.0  &   \nodata                      &    1.76$\pm$ 0.02  &     883  &    25.0  &    2324  &     -2.21  &   -1.81  &       $>$      72   \tablenotemark{h}  &  272  &  5.2  & 45.4 \\
  IRBG8  &14:15:38.1  & 52:18:52.9  &   \nodata                      &           \nodata  &    1217  &    24.9  &    2967  &   \nodata  &   -1.29  &                90             $\pm$ 6  &  122  &  0.1  & 45.7 \\
  IRBG9  &14:16:35.4  & 52:12:35.2  &   \nodata                      &           \nodata  &    2299  &    23.7  &    1999  &     -2.53  &   -1.92  &                92             $\pm$ 4  &  327  &  0.2  & 45.9 \\
 IRBG10  &14:18:51.5  & 52:48:35.9  &      1.92                      &    1.99$\pm$ 0.02  &    1686  &    24.0  &    1812  &   \nodata  &   -2.32  &                88             $\pm$ 3  &  394  &  0.3  & 45.6 \\
 IRBG11  &14:20:08.6  & 53:01:10.1  &   \nodata                      &    2.07$\pm$ 0.02  &     701  &    24.7  &    1491  &   \nodata  &   -1.94  &                92             $\pm$ 9  &   57  &  5.0  & 45.5 \\
 IRBG12  &14:20:13.0  & 52:55:33.3  &   \nodata                      &    1.04$\pm$ 0.01  &     948  &    24.8  &    2106  &   \nodata  &   -2.64  &                78             $\pm$ 3  &  218  &  0.1  & 44.3 \\
 IRBG13  &14:20:21.9  & 52:55:12.1  &   \nodata                      &    2.60$\pm$ 0.06  &    1786  &    24.1  &    2152  &   \nodata  &   -0.58  &                79             $\pm$ 5  &  440  &  1.4  & 46.1 \\
\enddata
\tablerefs{Spectroscopic redshifts from \citet[][IRBG4]{szokoly04}, \citet[][IRBG5, IRBG6]{swinbank04}, and \citet[][IRBG10]{davis07}}
\tablenotetext{a}{Optical/NIR spectroscopic redshift from the literature}
\tablenotetext{b}{IRS-derived redshift (this work); see \S4}
\tablenotetext{c}{IRAC (3.6-8.0 $\mu$m) and IRS (14-37 $\mu$m) power-law slopes, where $f_{\rm \nu} \propto \nu^{\alpha}$}
\tablenotetext{d}{AGN contribution to the MIR (14-37 \micron) flux density}
\tablenotetext{e}{Star-forming contribution to the observed 24 $\mu$m flux, as determined by the spectral fit}
\tablenotetext{f}{Optical depth of best-fit AGN continuum at 9.7$\mu$m (see \S4)}
\tablenotetext{g}{Rest-frame 6 \micron\ luminosity of the AGN compononent (see \S5.2)}
\tablenotetext{h}{IRS spectrum not fit by default model.  See \S4.1 for details.}
\end{deluxetable*}
\end{centering}

%% file: table3.tex
\begin{centering}
\begin{deluxetable}{cccccccccc}
\centering
\tabletypesize{\footnotesize}
\tablewidth{0pt}
\tablecaption {IR Selection Criteria}
\tablehead{
\colhead{ID}                     &
\colhead{H05}                     &
\colhead{Y05}                     &
\colhead{W06}                     &
\colhead{D08}                     &
\colhead{F08}                     &
\colhead{G08}                     &
\colhead{P08}                     &
\colhead{L09}                     &
\colhead{PL}               
}
\startdata
  IRBG1  &         x  &   \nodata  &         x  &         x  &         x  &         x  &   \nodata  &   \nodata  &         x \\
  IRBG2  &         x  &   \nodata  &   \nodata  &         x  &         x  &         x  &         x  &   \nodata  &         x \\
  IRBG3  &         x  &   \nodata  &   \nodata  &         x  &       $-$  &         x  &   \nodata  &   \nodata  &         x \\
  IRBG4  &         x  &   \nodata  &         x  &         x  &         x  &         x  &         x  &   \nodata  &         x \\
  IRBG5  &   \nodata  &         x  &   \nodata  &         x  &       $-$  &   \nodata  &         x  &   \nodata  &         x \\
  IRBG6  &   \nodata  &   \nodata  &   \nodata  &         x  &         x  &         x  &         x  &   \nodata  &         x \\
  IRBG7  &         x  &   \nodata  &   \nodata  &         x  &       $-$  &         x  &         x  &   \nodata  &         x \\
  IRBG8  &         x  &   \nodata  &         x  &         x  &         x  &         x  &         x  &   \nodata  &   \nodata \\
  IRBG9  &   \nodata  &   \nodata  &   \nodata  &         x  &         x  &         x  &         x  &   \nodata  &         x \\
 IRBG10  &   \nodata  &         x  &   \nodata  &         x  &   \nodata  &   \nodata  &   \nodata  &   \nodata  &   \nodata \\
 IRBG11  &   \nodata  &         x  &   \nodata  &         x  &   \nodata  &   \nodata  &   \nodata  &   \nodata  &   \nodata \\
 IRBG12  &         x  &         x  &   \nodata  &         x  &         x  &         x  &   \nodata  &   \nodata  &   \nodata \\
 IRBG13  &   \nodata  &   \nodata  &         x  &         x  &   \nodata  &   \nodata  &         x  &         x  &   \nodata \\
\enddata
\tablecomments{An 'x' indicates that a source would be selected via the listed criteria and a '$-$' indicates that a source lacked the neccessary data to determine if it would be selected.}
\tablerefs{H05: \cite{houck05}, Y05: \cite{yan05}, W06: \cite{weedman06red,weedman06irs}, D08: \cite{dey08} (DOGS), F08: \cite{fiore08}, G08: \cite{georgantopoulos08}, P08: \cite{polletta08}, L09: \cite{lanzuisi09} (EDOGS), PL (power-law): \cite{aah06,donley07}}
\end{deluxetable}
\end{centering}

%% file: table4.tex
\begin{centering}
\begin{deluxetable*}{lllllccc}
\centering
\tabletypesize{\footnotesize}
\tablewidth{0pt}
\tablecaption {X-ray Properties}
\tablehead{
\colhead{ID}                     &
\colhead{$T_{\rm X}$}             &
\colhead{$f_{(\rm{0.5-2\ keV})}$}             &
\colhead{$f_{(\rm{2-8\ keV})}$}             &
\colhead{$f_{(\rm{0.5-8\ keV})}$}             &
\colhead{log $L_{\rm{2-10\ keV}}$\tablenotemark{a}}   &
\colhead{log $N_{\rm H}$\tablenotemark{b}}  &
\colhead{log $L_{\rm{2-10\ keV, corr}}$\tablenotemark{c}}  \\ 
\colhead{}   &
\colhead{(s)}   &
\colhead{(ergs s$^{-1}$ cm$^{-2}$)}   &
\colhead{(ergs s$^{-1}$ cm$^{-2}$)}   &
\colhead{(ergs s$^{-1}$ cm$^{-2}$)}   &
\colhead{(cm$^{-2}$)}   &
\colhead{(ergs s$^{-1}$)}   &
\colhead{(ergs s$^{-1}$)}   
}
\startdata
  IRBG1  &  4.30E+05                      &    \hspace*{0.22cm}8.52E-16                      &    \hspace*{0.22cm}5.05E-15                      &    \hspace*{0.22cm}6.29E-15                      &    \hspace*{0.22cm}    43.6  &    \hspace*{0.22cm}      23.4    $^{+0.3}_{-0.3}$  &    \hspace*{0.22cm}      44.3 \\
  IRBG2  &  1.57E+06                      &    \hspace*{0.22cm}2.86E-15                      &    \hspace*{0.22cm}7.50E-15                      &    \hspace*{0.22cm}1.06E-14                      &    \hspace*{0.22cm}    44.2  &    \hspace*{0.22cm}      23.0    $^{+0.1}_{-0.1}$  &    \hspace*{0.22cm}      44.6 \\
  IRBG3  &  2.23E+05                      &                 $<$7.07E-17                      &    \hspace*{0.22cm}1.37E-15   \tablenotemark{d}  &    \hspace*{0.22cm}6.53E-16   \tablenotemark{d}  &                 $<$    42.3  &                 $>$      23.6                      &                 $>$      43.6 \\
  IRBG4  &  1.67E+06                      &    \hspace*{0.22cm}5.58E-15                      &    \hspace*{0.22cm}7.42E-15                      &    \hspace*{0.22cm}1.30E-14                      &    \hspace*{0.22cm}    44.2  &    \hspace*{0.22cm}      22.0    $^{+0.1}_{-0.1}$  &    \hspace*{0.22cm}      44.2 \\
  IRBG5  &  1.65E+06                      &                 $<$4.00E-17                      &                 $<$3.26E-16                      &                 $<$2.06E-16                      &                 $<$    42.3  &    \hspace*{0.22cm}   \nodata                      &    \hspace*{0.22cm}   \nodata \\
  IRBG6  &  1.92E+06                      &    \hspace*{0.22cm}2.07E-16                      &    \hspace*{0.22cm}2.48E-15                      &    \hspace*{0.22cm}2.52E-15                      &    \hspace*{0.22cm}    43.0  &    \hspace*{0.22cm}      23.6    $^{+0.0}_{-0.1}$  &    \hspace*{0.22cm}      44.0 \\
  IRBG7  &  1.41E+06                      &    \hspace*{0.22cm}1.52E-16                      &    \hspace*{0.22cm}1.32E-15                      &    \hspace*{0.22cm}1.40E-15                      &    \hspace*{0.22cm}    42.7  &    \hspace*{0.22cm}      23.4    $^{+0.1}_{-0.2}$  &    \hspace*{0.22cm}      43.6 \\
  IRBG8  &  7.16E+07   \tablenotemark{e}  &    \hspace*{0.22cm}4.28E-15                      &    \hspace*{0.22cm}1.04E-14                      &    \hspace*{0.22cm}1.47E-14                      &    \hspace*{0.22cm}    44.3  &    \hspace*{0.22cm}      22.8    $^{+0.3}_{-0.3}$  &    \hspace*{0.22cm}      44.6 \\
  IRBG9  &  4.86E+07   \tablenotemark{e}  &                 $<$1.86E-16                      &                 $<$1.10E-15                      &    \hspace*{0.22cm}5.39E-16   \tablenotemark{d}  &                 $<$    42.9  &    \hspace*{0.22cm}   \nodata                      &    \hspace*{0.22cm}   \nodata \\
 IRBG10  &  5.10E+07   \tablenotemark{e}  &    \hspace*{0.22cm}1.86E-16                      &                 $<$4.85E-16                      &    \hspace*{0.22cm}5.11E-16                      &    \hspace*{0.22cm}    42.9  &                 $<$      22.8                      &                 $<$      43.2 \\
 IRBG11  &  5.78E+07   \tablenotemark{e}  &                 $<$6.64E-17                      &                 $<$8.63E-17                      &                 $<$1.53E-16                      &                 $<$    42.5  &    \hspace*{0.22cm}   \nodata                      &    \hspace*{0.22cm}   \nodata \\
 IRBG12  &  7.34E+07   \tablenotemark{e}  &                 $<$1.74E-16                      &                 $<$7.76E-16                      &                 $<$6.49E-16                      &                 $<$    42.3  &    \hspace*{0.22cm}   \nodata                      &    \hspace*{0.22cm}   \nodata \\
 IRBG13  &  7.47E+07   \tablenotemark{e}  &                 $<$1.24E-16                      &                 $<$7.95E-16                      &                 $<$4.93E-16                      &                 $<$    43.0  &    \hspace*{0.22cm}   \nodata                      &    \hspace*{0.22cm}   \nodata \\
\enddata
\tablenotetext{a}{\ Log of the rest-frame, non absorption-corrected 2-10 keV luminosity, calculated using the observed 0.5-2 keV flux, $\Gamma = 2.0$, and an assumed redshift of $z=2.0$ for those sources lacking redshift constraints.  Errors are based on the uncertainties in the soft-band flux.}
\tablenotetext{b}{Estimated using the observed flux ratio and an intrinsic photon index of $\Gamma = 2$}
\tablenotetext{c}{\ Log of the rest-frame, absorption-corrected 2-10 keV luminosity}
\tablenotetext{d}{\ Low-significance X-ray detections: $\sigma$(IRBG3, 0.5-8 keV) = 4.8; $\sigma$(IRBG3, 2-8 keV) = 5.5; $\sigma$(IRBG9, 0.5-8 keV) = 2.1}
\tablenotetext{e}{Units of s cm$^2$, nominal exposure time is 200~ks}
\end{deluxetable*}
\end{centering}

%% file: ms.bbl
\begin{thebibliography}{}

\bibitem[{{Alexander} {et~al.}(2003){Alexander}, {Bauer}, {Brandt},
  {Schneider}, {Hornschemeier}, {Vignali}, {Barger}, {Broos}, {Cowie},
  {Garmire}, {Townsley}, {Bautz}, {Chartas}, \& {Sargent}}]{alexander03}
{Alexander}, D.~M., {et~al.} 2003, \aj, 126, 539

\bibitem[{{Alexander} {et~al.}(2005){Alexander}, {Bauer}, {Chapman}, {Smail},
  {Blain}, {Brandt}, \& {Ivison}}]{alexander05}
{Alexander}, D.~M., {Bauer}, F.~E., {Chapman}, S.~C., {Smail}, I., {Blain},
  A.~W., {Brandt}, W.~N., \& {Ivison}, R.~J. 2005, \apj, 632, 736

\bibitem[{{Alexander} {et~al.}(2008{\natexlab{a}}){Alexander}, {Brandt},
  {Smail}, {Swinbank}, {Bauer}, {Blain}, {Chapman}, {Coppin}, {Ivison}, \&
  {Men{\'e}ndez-Delmestre}}]{alexander08smg}
{Alexander}, D.~M., {et~al.} 2008{\natexlab{a}}, \aj, 135, 1968

\bibitem[{{Alexander} {et~al.}(2008{\natexlab{b}}){Alexander}, {Chary}, {Pope},
  {Bauer}, {Brandt}, {Daddi}, {Dickinson}, {Elbaz}, \& {Reddy}}]{alexander08ct}
---. 2008{\natexlab{b}}, \apj, 687, 835

\bibitem[{{Alonso-Herrero} {et~al.}(2006){Alonso-Herrero},
  {P{\'e}rez-Gonz{\'a}lez}, {Alexander}, {Rieke}, {Rigopoulou}, {Le Floc'h},
  {Barmby}, {Papovich}, {Rigby}, {Bauer}, {Brandt}, {Egami}, {Willner}, {Dole},
  \& {Huang}}]{aah06}
{Alonso-Herrero}, A., {et~al.} 2006, \apj, 640, 167

\bibitem[{{Arnouts} {et~al.}(2002){Arnouts}, {Vandame}, {Benoist},
  {Groenewegen}, {da Costa}, {Schirmer}, {Mignani}, \& {Slijkhuis}}]{arnouts02}
{Arnouts}, S., {Vandame}, B., {Benoist}, C., {Groenewegen}, M.~A.~T., {da
  Costa}, L., {Schirmer}, M., {Mignani}, R.~P., \& {Slijkhuis}, R. 2002, VizieR
  Online Data Catalog, 337, 90740

\bibitem[{{Assef} {et~al.}(2008){Assef}, {Kochanek}, {Brodwin}, {Brown},
  {Caldwell}, {Cool}, {Eisenhardt}, {Eisenstein}, {Gonzalez}, {Jannuzi},
  {Jones}, {McKenzie}, {Murray}, \& {Stern}}]{assef08}
{Assef}, R.~J., {et~al.} 2008, \apj, 676, 286

\bibitem[{{Beswick} {et~al.}(2008){Beswick}, {Muxlow}, {Thrall}, {Richards}, \&
  {Garrington}}]{beswick08}
{Beswick}, R.~J., {Muxlow}, T.~W.~B., {Thrall}, H., {Richards}, A.~M.~S., \&
  {Garrington}, S.~T. 2008, \mnras, 385, 1143

\bibitem[{{Blain} {et~al.}(2004){Blain}, {Chapman}, {Smail}, \&
  {Ivison}}]{blain04}
{Blain}, A.~W., {Chapman}, S.~C., {Smail}, I., \& {Ivison}, R. 2004, \apj, 611,
  725

\bibitem[{{Boyle} {et~al.}(2007){Boyle}, {Cornwell}, {Middelberg}, {Norris},
  {Appleton}, \& {Smail}}]{boyle07}
{Boyle}, B.~J., {Cornwell}, T.~J., {Middelberg}, E., {Norris}, R.~P.,
  {Appleton}, P.~N., \& {Smail}, I. 2007, \mnras, 376, 1182

\bibitem[{{Brand} {et~al.}(2007){Brand}, {Dey}, {Desai}, {Soifer}, {Bian},
  {Armus}, {Brown}, {Le Floc'h}, {Higdon}, {Houck}, {Jannuzi}, \&
  {Weedman}}]{brand07}
{Brand}, K., {et~al.} 2007, \apj, 663, 204

\bibitem[{{Brand} {et~al.}(2008){Brand}, {Weedman}, {Desai}, {Le Floc'h},
  {Armus}, {Dey}, {Houck}, {Jannuzi}, {Smith}, \& {Soifer}}]{brand08}
---. 2008, \apj, 680, 119

\bibitem[{{Brodwin} {et~al.}(2008){Brodwin}, {Dey}, {Brown}, {Pope}, {Armus},
  {Bussmann}, {Desai}, {Jannuzi}, \& {Le Floc'h}}]{brodwin08}
{Brodwin}, M., {et~al.} 2008, \apjl, 687, L65

\bibitem[{{Buchanan} {et~al.}(2006){Buchanan}, {Gallimore}, {O'Dea}, {Baum},
  {Axon}, {Robinson}, {Elitzur}, \& {Elvis}}]{buchanan06}
{Buchanan}, C.~L., {Gallimore}, J.~F., {O'Dea}, C.~P., {Baum}, S.~A., {Axon},
  D.~J., {Robinson}, A., {Elitzur}, M., \& {Elvis}, M. 2006, \aj, 132, 401

\bibitem[{{Bussmann} {et~al.}(2009){Bussmann}, {Dey}, {Lotz}, {Armus}, {Brand},
  {Brown}, {Desai}, {Eisenhardt}, {Higdon}, {Higdon}, {Jannuzi}, {LeFloc'h},
  {Melbourne}, {Soifer}, \& {Weedman}}]{bussmann09}
{Bussmann}, R.~S., {et~al.} 2009, \apj, 693, 750

\bibitem[{{Capak} {et~al.}(2004){Capak}, {Cowie}, {Hu}, {Barger}, {Dickinson},
  {Fernandez}, {Giavalisco}, {Komiyama}, {Kretchmer}, {McNally}, {Miyazaki},
  {Okamura}, \& {Stern}}]{capak04}
{Capak}, P., {et~al.} 2004, \aj, 127, 180

\bibitem[{{Chapman} {et~al.}(2005){Chapman}, {Blain}, {Smail}, \&
  {Ivison}}]{chapman05}
{Chapman}, S.~C., {Blain}, A.~W., {Smail}, I., \& {Ivison}, R.~J. 2005, \apj,
  622, 772

\bibitem[{{Conselice} {et~al.}(2003{\natexlab{a}}){Conselice}, {Bershady},
  {Dickinson}, \& {Papovich}}]{conselice03mergers}
{Conselice}, C.~J., {Bershady}, M.~A., {Dickinson}, M., \& {Papovich}, C.
  2003{\natexlab{a}}, \aj, 126, 1183

\bibitem[{{Conselice} {et~al.}(2000){Conselice}, {Bershady}, \&
  {Jangren}}]{conselice00}
{Conselice}, C.~J., {Bershady}, M.~A., \& {Jangren}, A. 2000, \apj, 529, 886

\bibitem[{{Conselice} {et~al.}(2005){Conselice}, {Blackburne}, \&
  {Papovich}}]{conselice05}
{Conselice}, C.~J., {Blackburne}, J.~A., \& {Papovich}, C. 2005, \apj, 620, 564

\bibitem[{{Conselice} {et~al.}(2003{\natexlab{b}}){Conselice}, {Chapman}, \&
  {Windhorst}}]{conselice03smg}
{Conselice}, C.~J., {Chapman}, S.~C., \& {Windhorst}, R.~A. 2003{\natexlab{b}},
  \apjl, 596, L5

\bibitem[{{Coppin} {et~al.}(2006){Coppin}, {Chapin}, {Mortier}, {Scott},
  {Borys}, {Dunlop}, {Halpern}, {Hughes}, {Pope}, {Scott}, {Serjeant}, {Wagg},
  {Alexander}, {Almaini}, {Aretxaga}, {Babbedge}, {Best}, {Blain}, {Chapman},
  {Clements}, {Crawford}, {Dunne}, {Eales}, {Edge}, {Farrah}, {Gazta{\~n}aga},
  {Gear}, {Granato}, {Greve}, {Fox}, {Ivison}, {Jarvis}, {Jenness}, {Lacey},
  {Lepage}, {Mann}, {Marsden}, {Martinez-Sansigre}, {Oliver}, {Page},
  {Peacock}, {Pearson}, {Percival}, {Priddey}, {Rawlings}, {Rowan-Robinson},
  {Savage}, {Seigar}, {Sekiguchi}, {Silva}, {Simpson}, {Smail}, {Stevens},
  {Takagi}, {Vaccari}, {van Kampen}, \& {Willott}}]{coppin06}
{Coppin}, K., {et~al.} 2006, \mnras, 372, 1621

\bibitem[{{Coppin} {et~al.}(2010){Coppin}, {Pope}, {Menendez-Delmestre},
  {Alexander}, {Dunlop}, {Egami}, {Gabor}, {Ibar}, {Ivison}, {Austermann},
  {Blain}, {Chapman}, {Clements}, {Dunne}, {Dye}, {Farrah}, {Hughes},
  {Mortier}, {Page}, {Rowan-Robinson}, {Scott}, {Simpson}, {Smail}, {Swinbank},
  {Vaccari}, \& {Yun}}]{coppin10}
---. 2010, ArXiv e-prints

\bibitem[{{Cowie} {et~al.}(2004){Cowie}, {Barger}, {Hu}, {Capak}, \&
  {Songaila}}]{cowie04}
{Cowie}, L.~L., {Barger}, A.~J., {Hu}, E.~M., {Capak}, P., \& {Songaila}, A.
  2004, \aj, 127, 3137

\bibitem[{{Dasyra} {et~al.}(2008){Dasyra}, {Yan}, {Helou}, {Surace}, {Sajina},
  \& {Colbert}}]{dasyra08}
{Dasyra}, K.~M., {Yan}, L., {Helou}, G., {Surace}, J., {Sajina}, A., \&
  {Colbert}, J. 2008, \apj, 680, 232

\bibitem[{{Davis} {et~al.}(2007){Davis}, {Guhathakurta}, {Konidaris}, {Newman},
  {Ashby}, {Biggs}, {Barmby}, {Bundy}, {Chapman}, {Coil}, {Conselice},
  {Cooper}, {Croton}, {Eisenhardt}, {Ellis}, {Faber}, {Fang}, {Fazio},
  {Georgakakis}, {Gerke}, {Goss}, {Gwyn}, {Harker}, {Hopkins}, {Huang},
  {Ivison}, {Kassin}, {Kirby}, {Koekemoer}, {Koo}, {Laird}, {Le Floc'h}, {Lin},
  {Lotz}, {Marshall}, {Martin}, {Metevier}, {Moustakas}, {Nandra}, {Noeske},
  {Papovich}, {Phillips}, {Rich}, {Rieke}, {Rigopoulou}, {Salim},
  {Schiminovich}, {Simard}, {Smail}, {Small}, {Weiner}, {Willmer}, {Willner},
  {Wilson}, {Wright}, \& {Yan}}]{davis07}
{Davis}, M., {et~al.} 2007, \apjl, 660, L1

\bibitem[{{Desai} {et~al.}(2009){Desai}, {Soifer}, {Dey}, {Le Floc'h}, {Armus},
  {Brand}, {Brown}, {Brodwin}, {Jannuzi}, {Houck}, {Weedman}, {Ashby},
  {Gonzalez}, {Huang}, {Smith}, {Teplitz}, {Willner}, \& {Melbourne}}]{desai09}
{Desai}, V., {et~al.} 2009, \apj, 700, 1190

\bibitem[{{Dey} {et~al.}(2008){Dey}, {Soifer}, {Desai}, {Brand}, {Le Floc'h},
  {Brown}, {Jannuzi}, {Armus}, {Bussmann}, {Brodwin}, {Bian}, {Eisenhardt},
  {Higdon}, {Weedman}, \& {Willner}}]{dey08}
{Dey}, A., {et~al.} 2008, \apj, 677, 943

\bibitem[{{Donley} {et~al.}(2008){Donley}, {Rieke}, {P{\'e}rez-Gonz{\'a}lez},
  \& {Barro}}]{donley08}
{Donley}, J.~L., {Rieke}, G.~H., {P{\'e}rez-Gonz{\'a}lez}, P.~G., \& {Barro},
  G. 2008, \apj, 687, 111

\bibitem[{{Donley} {et~al.}(2007){Donley}, {Rieke}, {P{\'e}rez-Gonz{\'a}lez},
  {Rigby}, \& {Alonso-Herrero}}]{donley07}
{Donley}, J.~L., {Rieke}, G.~H., {P{\'e}rez-Gonz{\'a}lez}, P.~G., {Rigby},
  J.~R., \& {Alonso-Herrero}, A. 2007, \apj, 660, 167

\bibitem[{{Donley} {et~al.}(2005){Donley}, {Rieke}, {Rigby}, \&
  {P{\'e}rez-Gonz{\'a}lez}}]{donley05}
{Donley}, J.~L., {Rieke}, G.~H., {Rigby}, J.~R., \& {P{\'e}rez-Gonz{\'a}lez},
  P.~G. 2005, \apj, 634, 169

\bibitem[{{Draine}(2003)}]{draine03}
{Draine}, B.~T. 2003, \araa, 41, 241

\bibitem[{{Efstathiou} {et~al.}(1995){Efstathiou}, {Hough}, \&
  {Young}}]{efstathiou95}
{Efstathiou}, A., {Hough}, J.~H., \& {Young}, S. 1995, \mnras, 277, 1134


\bibitem[{Fadda} {et~al.}(2010)]{fadda10}
{Fadda}, D., et al. 2010, \apj, submitted

\bibitem[{{Farrah} {et~al.}(2008){Farrah}, {Lonsdale}, {Weedman}, {Spoon},
  {Rowan-Robinson}, {Polletta}, {Oliver}, {Houck}, \& {Smith}}]{farrah08}
{Farrah}, D., {et~al.} 2008, \apj, 677, 957

\bibitem[{{Fiore} {et~al.}(2008){Fiore}, {Grazian}, {Santini}, {Puccetti},
  {Brusa}, {Feruglio}, {Fontana}, {Giallongo}, {Comastri}, {Gruppioni},
  {Pozzi}, {Zamorani}, \& {Vignali}}]{fiore08}
{Fiore}, F., {et~al.} 2008, \apj, 672, 94

\bibitem[{{Fiore} {et~al.}(2009){Fiore}, {Puccetti}, {Brusa}, {Salvato},
  {Zamorani}, {Aldcroft}, {Aussel}, {Brunner}, {Capak}, {Cappelluti}, {Civano},
  {Comastri}, {Elvis}, {Feruglio}, {Finoguenov}, {Fruscione}, {Gilli},
  {Hasinger}, {Koekemoer}, {Kartaltepe}, {Ilbert}, {Impey}, {LeFloc'h},
  {Lilly}, {Mainieri}, {Martinez-Sansigre}, {McCracken}, {Menci}, {Merloni},
  {Miyaji}, {Sanders}, {Sargent}, {Schinnerer}, {Scoville}, {Silverman},
  {Smolcic}, {Steffen}, {Santini}, {Taniguchi}, {Thompson}, {Trump}, {Vignali},
  {Urry}, \& {Yan}}]{fiore09}
---. 2009, \apj, 693, 447

\bibitem[{{Franceschini} {et~al.}(2003){Franceschini}, {Braito}, {Persic},
  {Della Ceca}, {Bassani}, {Cappi}, {Malaguti}, {Palumbo}, {Risaliti},
  {Salvati}, \& {Severgnini}}]{franceschini03}
{Franceschini}, A., {et~al.} 2003, \mnras, 343, 1181

\bibitem[{{Gendreau} {et~al.}(1995){Gendreau}, {Mushotzky}, {Fabian}, {Holt},
  {Kii}, {Serlemitsos}, {Ogasaka}, {Tanaka}, {Bautz}, {Fukazawa}, {Ishisaki},
  {Kohmura}, {Makishima}, {Tashiro}, {Tsusaka}, {Kunieda}, {Ricker}, \&
  {Vanderspek}}]{gendreau95}
{Gendreau}, K.~C., {et~al.} 1995, \pasj, 47, L5

\bibitem[{{Georgantopoulos} {et~al.}(2009){Georgantopoulos}, {Akylas},
  {Georgakakis}, \& {Rowan-Robinson}}]{georgantopoulos09}
{Georgantopoulos}, I., {Akylas}, A., {Georgakakis}, A., \& {Rowan-Robinson}, M.
  2009, \aap, 507, 747

\bibitem[{{Georgantopoulos} {et~al.}(2008){Georgantopoulos}, {Georgakakis},
  {Rowan-Robinson}, \& {Rovilos}}]{georgantopoulos08}
{Georgantopoulos}, I., {Georgakakis}, A., {Rowan-Robinson}, M., \& {Rovilos},
  E. 2008, \aap, 484, 671

\bibitem[{{George} {et~al.}(2000){George}, {Turner}, {Yaqoob}, {Netzer},
  {Laor}, {Mushotzky}, {Nandra}, \& {Takahashi}}]{george00}
{George}, I.~M., {Turner}, T.~J., {Yaqoob}, T., {Netzer}, H., {Laor}, A.,
  {Mushotzky}, R.~F., {Nandra}, K., \& {Takahashi}, T. 2000, \apj, 531, 52

\bibitem[{{Giavalisco} {et~al.}(2004){Giavalisco}, {Ferguson}, {Koekemoer},
  {Dickinson}, {Alexander}, {Bauer}, {Bergeron}, {Biagetti}, {Brandt},
  {Casertano}, {Cesarsky}, {Chatzichristou}, {Conselice}, {Cristiani}, {Da
  Costa}, {Dahlen}, {de Mello}, {Eisenhardt}, {Erben}, {Fall}, {Fassnacht},
  {Fosbury}, {Fruchter}, {Gardner}, {Grogin}, {Hook}, {Hornschemeier}, {Idzi},
  {Jogee}, {Kretchmer}, {Laidler}, {Lee}, {Livio}, {Lucas}, {Madau},
  {Mobasher}, {Moustakas}, {Nonino}, {Padovani}, {Papovich}, {Park},
  {Ravindranath}, {Renzini}, {Richardson}, {Riess}, {Rosati}, {Schirmer},
  {Schreier}, {Somerville}, {Spinrad}, {Stern}, {Stiavelli}, {Strolger},
  {Urry}, {Vandame}, {Williams}, \& {Wolf}}]{giavalisco04}
{Giavalisco}, M., {et~al.} 2004, \apjl, 600, L93

\bibitem[{{Hainline} {et~al.}(2009){Hainline}, {Blain}, {Smail}, {Frayer},
  {Chapman}, {Ivison}, \& {Alexander}}]{hainline09}
{Hainline}, L.~J., {Blain}, A.~W., {Smail}, I., {Frayer}, D.~T., {Chapman},
  S.~C., {Ivison}, R.~J., \& {Alexander}, D.~M. 2009, \apj, 699, 1610

\bibitem[{{Heckman}(1995)}]{heckman95}
{Heckman}, T.~M. 1995, \apj, 446, 101

\bibitem[{{Hickox} \& {Markevitch}(2006)}]{hickox06}
{Hickox}, R.~C., \& {Markevitch}, M. 2006, \apj, 645, 95

\bibitem[{{Hopkins} {et~al.}(2008){Hopkins}, {Hernquist}, {Cox}, \& {Kere{\v
  s}}}]{hopkins08}
{Hopkins}, P.~F., {Hernquist}, L., {Cox}, T.~J., \& {Kere{\v s}}, D. 2008,
  \apjs, 175, 356

\bibitem[{{Houck} {et~al.}(2004){Houck}, {Roellig}, {van Cleve}, {Forrest},
  {Herter}, {Lawrence}, {Matthews}, {Reitsema}, {Soifer}, {Watson}, {Weedman},
  {Huisjen}, {Troeltzsch}, {Barry}, {Bernard-Salas}, {Blacken}, {Brandl},
  {Charmandaris}, {Devost}, {Gull}, {Hall}, {Henderson}, {Higdon}, {Pirger},
  {Schoenwald}, {Sloan}, {Uchida}, {Appleton}, {Armus}, {Burgdorf},
  {Fajardo-Acosta}, {Grillmair}, {Ingalls}, {Morris}, \& {Teplitz}}]{houck04}
{Houck}, J.~R., {et~al.} 2004, \apjs, 154, 18

\bibitem[{{Houck} {et~al.}(2005){Houck}, {Soifer}, {Weedman}, {Higdon},
  {Higdon}, {Herter}, {Brown}, {Dey}, {Jannuzi}, {Le Floc'h}, {Rieke}, {Armus},
  {Charmandaris}, {Brandl}, \& {Teplitz}}]{houck05}
---. 2005, \apjl, 622, L105

\bibitem[{{Ibar} {et~al.}(2008){Ibar}, {Cirasuolo}, {Ivison}, {Best}, {Smail},
  {Biggs}, {Simpson}, {Dunlop}, {Almaini}, {McLure}, {Foucaud}, \&
  {Rawlings}}]{ibar08}
{Ibar}, E., {et~al.} 2008, \mnras, 386, 953

\bibitem[{{Ivison} {et~al.}(2007){Ivison}, {Chapman}, {Faber}, {Smail},
  {Biggs}, {Conselice}, {Wilson}, {Salim}, {Huang}, \& {Willner}}]{ivison07}
{Ivison}, R.~J., {et~al.} 2007, \apjl, 660, L77

\bibitem[{{Kellermann} {et~al.}(2008){Kellermann}, {Fomalont}, {Mainieri},
  {Padovani}, {Rosati}, {Shaver}, {Tozzi}, \& {Miller}}]{kellermann08}
{Kellermann}, K.~I., {Fomalont}, E.~B., {Mainieri}, V., {Padovani}, P.,
  {Rosati}, P., {Shaver}, P., {Tozzi}, P., \& {Miller}, N. 2008, \apjs, 179, 71

\bibitem[{{Laird} {et~al.}(2009){Laird}, {Nandra}, {Georgakakis}, {Aird},
  {Barmby}, {Conselice}, {Coil}, {Davis}, {Faber}, {Fazio}, {Guhathakurta},
  {Koo}, {Sarajedini}, \& {Willmer}}]{laird09}
{Laird}, E.~S., {et~al.} 2009, \apjs, 180, 102

\bibitem[{{Lanzuisi} {et~al.}(2009){Lanzuisi}, {Piconcelli}, {Fiore},
  {Feruglio}, {Vignali}, {Salvato}, \& {Gruppioni}}]{lanzuisi09}
{Lanzuisi}, G., {Piconcelli}, E., {Fiore}, F., {Feruglio}, C., {Vignali}, C.,
  {Salvato}, M., \& {Gruppioni}, C. 2009, ArXiv e-prints

\bibitem[{{Le F{\`e}vre} {et~al.}(2004){Le F{\`e}vre}, {Mellier}, {McCracken},
  {Foucaud}, {Gwyn}, {Radovich}, {Dantel-Fort}, {Bertin}, {Moreau},
  {Cuillandre}, {Pierre}, {Le Brun}, {Mazure}, \& {Tresse}}]{lefevre04}
{Le F{\`e}vre}, O., {et~al.} 2004, \aap, 417, 839

\bibitem[{{Lehmer} {et~al.}(2005){Lehmer}, {Brandt}, {Alexander}, {Bauer},
  {Schneider}, {Tozzi}, {Bergeron}, {Garmire}, {Giacconi}, {Gilli}, {Hasinger},
  {Hornschemeier}, {Koekemoer}, {Mainieri}, {Miyaji}, {Nonino}, {Rosati},
  {Silverman}, {Szokoly}, \& {Vignali}}]{lehmer05}
{Lehmer}, B.~D., {et~al.} 2005, \apjs, 161, 21

\bibitem[{{Lotz} {et~al.}(2008){Lotz}, {Jonsson}, {Cox}, \& {Primack}}]{lotz08}
{Lotz}, J.~M., {Jonsson}, P., {Cox}, T.~J., \& {Primack}, J.~R. 2008, \mnras,
  391, 1137

\bibitem[{{Luo} {et~al.}(2008){Luo}, {Bauer}, {Brandt}, {Alexander}, {Lehmer},
  {Schneider}, {Brusa}, {Comastri}, {Fabian}, {Finoguenov}, {Gilli},
  {Hasinger}, {Hornschemeier}, {Koekemoer}, {Mainieri}, {Paolillo}, {Rosati},
  {Shemmer}, {Silverman}, {Smail}, {Steffen}, \& {Vignali}}]{luo08}
{Luo}, B., {et~al.} 2008, \apjs, 179, 19

\bibitem[{{Lutz} {et~al.}(2004){Lutz}, {Maiolino}, {Spoon}, \&
  {Moorwood}}]{lutz04}
{Lutz}, D., {Maiolino}, R., {Spoon}, H.~W.~W., \& {Moorwood}, A.~F.~M. 2004,
  \aap, 418, 465

\bibitem[{{Maiolino} {et~al.}(2007){Maiolino}, {Shemmer}, {Imanishi}, {Netzer},
  {Oliva}, {Lutz}, \& {Sturm}}]{maiolino07}
{Maiolino}, R., {Shemmer}, O., {Imanishi}, M., {Netzer}, H., {Oliva}, E.,
  {Lutz}, D., \& {Sturm}, E. 2007, \aap, 468, 979

\bibitem[{{Marshall} {et~al.}(1980){Marshall}, {Boldt}, {Holt}, {Miller},
  {Mushotzky}, {Rose}, {Rothschild}, \& {Serlemitsos}}]{marshall80}
{Marshall}, F.~E., {Boldt}, E.~A., {Holt}, S.~S., {Miller}, R.~B., {Mushotzky},
  R.~F., {Rose}, L.~A., {Rothschild}, R.~E., \& {Serlemitsos}, P.~J. 1980,
  \apj, 235, 4

\bibitem[{{Marzke} {et~al.}(1999){Marzke}, {McCarthy}, {Persson}, {Oemler},
  {Dressler}, {Yan}, {Carlberg}, {Abraham}, {Ellis}, {Firth}, {Mackay}, \&
  {McMahon}}]{marzke99}
{Marzke}, R., {et~al.} 1999, in Astronomical Society of the Pacific Conference
  Series, Vol. 191, Photometric Redshifts and the Detection of High Redshift
  Galaxies, ed. R.~{Weymann}, L.~{Storrie-Lombardi}, M.~{Sawicki}, \&
  R.~{Brunner}, 148--+

\bibitem[{{Mason} {et~al.}(2006){Mason}, {Geballe}, {Packham}, {Levenson},
  {Elitzur}, {Fisher}, \& {Perlman}}]{mason06}
{Mason}, R.~E., {Geballe}, T.~R., {Packham}, C., {Levenson}, N.~A., {Elitzur},
  M., {Fisher}, R.~S., \& {Perlman}, E. 2006, \apj, 640, 612

\bibitem[{{McLure} {et~al.}(2006){McLure}, {Jarvis}, {Targett}, {Dunlop}, \&
  {Best}}]{mclure06}
{McLure}, R.~J., {Jarvis}, M.~J., {Targett}, T.~A., {Dunlop}, J.~S., \& {Best},
  P.~N. 2006, \mnras, 368, 1395

\bibitem[{{Melbourne} {et~al.}(2009){Melbourne}, {Bussman}, {Brand}, {Desai},
  {Armus}, {Dey}, {Jannuzi}, {Houck}, {Matthews}, \& {Soifer}}]{melbourne09}
{Melbourne}, J., {et~al.} 2009, \aj, 137, 4854

\bibitem[{{Men{\'e}ndez-Delmestre} {et~al.}(2009){Men{\'e}ndez-Delmestre},
  {Blain}, {Smail}, {Alexander}, {Chapman}, {Armus}, {Frayer}, {Ivison}, \&
  {Teplitz}}]{menendez09}
{Men{\'e}ndez-Delmestre}, K., {et~al.} 2009, \apj, 699, 667

\bibitem[{{Murphy} {et~al.}(2009){Murphy}, {Chary}, {Alexander}, {Dickinson},
  {Magnelli}, {Morrison}, {Pope}, \& {Teplitz}}]{murphy09}
{Murphy}, E.~J., {Chary}, R., {Alexander}, D.~M., {Dickinson}, M., {Magnelli},
  B., {Morrison}, G., {Pope}, A., \& {Teplitz}, H.~I. 2009, \apj, 698, 1380

\bibitem[{{Narayanan} {et~al.}(2009{\natexlab{a}}){Narayanan}, {Dey},
  {Hayward}, {Cox}, {Bussmann}, {Brodwin}, {Jonsson}, {Hopkins}, {Groves},
  {Younger}, \& {Hernquist}}]{desika09dog}
{Narayanan}, D., {et~al.} 2009{\natexlab{a}}, ArXiv e-prints

\bibitem[{{Narayanan} {et~al.}(2009{\natexlab{b}}){Narayanan}, {Hayward},
  {Cox}, {Hernquist}, {Jonsson}, {Younger}, \& {Groves}}]{desika09smg}
{Narayanan}, D., {Hayward}, C.~C., {Cox}, T.~J., {Hernquist}, L., {Jonsson},
  P., {Younger}, J.~D., \& {Groves}, B. 2009{\natexlab{b}}, \mnras, 1712

\bibitem[{{Papovich} {et~al.}(2007){Papovich}, {Rudnick}, {Le Floc'h}, {van
  Dokkum}, {Rieke}, {Taylor}, {Armus}, {Gawiser}, {Huang}, {Marcillac}, \&
  {Franx}}]{papovich07}
{Papovich}, C., {et~al.} 2007, \apj, 668, 45

\bibitem[{{Peng} {et~al.}(2006){Peng}, {Impey}, {Rix}, {Kochanek}, {Keeton},
  {Falco}, {Leh{\'a}r}, \& {McLeod}}]{peng06}
{Peng}, C.~Y., {Impey}, C.~D., {Rix}, H., {Kochanek}, C.~S., {Keeton}, C.~R.,
  {Falco}, E.~E., {Leh{\'a}r}, J., \& {McLeod}, B.~A. 2006, \apj, 649, 616

\bibitem[{{P{\'e}rez-Gonz{\'a}lez} {et~al.}(2005){P{\'e}rez-Gonz{\'a}lez},
  {Rieke}, {Egami}, {Alonso-Herrero}, {Dole}, {Papovich}, {Blaylock}, {Jones},
  {Rieke}, {Rigby}, {Barmby}, {Fazio}, {Huang}, \& {Martin}}]{pgperez05}
{P{\'e}rez-Gonz{\'a}lez}, P.~G., {et~al.} 2005, \apj, 630, 82

\bibitem[{{P{\'e}rez-Gonz{\'a}lez} {et~al.}(2008){P{\'e}rez-Gonz{\'a}lez},
  {Rieke}, {Villar}, {Barro}, {Blaylock}, {Egami}, {Gallego}, {Gil de Paz},
  {Pascual}, {Zamorano}, \& {Donley}}]{pgperez08}
---. 2008, \apj, 675, 234

\bibitem[{{Perola} {et~al.}(2004){Perola}, {Puccetti}, {Fiore}, {Sacchi},
  {Brusa}, {Cocchia}, {Baldi}, {Carangelo}, {Ciliegi}, {Comastri}, {La Franca},
  {Maiolino}, {Matt}, {Mignoli}, {Molendi}, \& {Vignali}}]{perola04}
{Perola}, G.~C., {et~al.} 2004, \aap, 421, 491

\bibitem[{{Pier} \& {Krolik}(1992)}]{pier92}
{Pier}, E.~A., \& {Krolik}, J.~H. 1992, \apj, 401, 99

\bibitem[{{Polletta} {et~al.}(2008){Polletta}, {Weedman}, {H{\"o}nig},
  {Lonsdale}, {Smith}, \& {Houck}}]{polletta08}
{Polletta}, M., {Weedman}, D., {H{\"o}nig}, S., {Lonsdale}, C.~J., {Smith},
  H.~E., \& {Houck}, J. 2008, \apj, 675, 960

\bibitem[{{Polletta} {et~al.}(2006){Polletta}, {Wilkes}, {Siana}, {Lonsdale},
  {Kilgard}, {Smith}, {Kim}, {Owen}, {Efstathiou}, {Jarrett}, {Stacey},
  {Franceschini}, {Rowan-Robinson}, {Babbedge}, {Berta}, {Fang}, {Farrah},
  {Gonz{\'a}lez-Solares}, {Morrison}, {Surace}, \& {Shupe}}]{polletta06}
{Polletta}, M.~d.~C., {et~al.} 2006, \apj, 642, 673

\bibitem[{{Pope} {et~al.}(2005){Pope}, {Borys}, {Scott}, {Conselice},
  {Dickinson}, \& {Mobasher}}]{pope05}
{Pope}, A., {Borys}, C., {Scott}, D., {Conselice}, C., {Dickinson}, M., \&
  {Mobasher}, B. 2005, \mnras, 358, 149

\bibitem[{{Pope} {et~al.}(2008{\natexlab{a}}){Pope}, {Bussmann}, {Dey},
  {Meger}, {Alexander}, {Brodwin}, {Chary}, {Dickinson}, {Frayer}, {Greve},
  {Huynh}, {Lin}, {Morrison}, {Scott}, \& {Yan}}]{pope08dog}
{Pope}, A., {et~al.} 2008{\natexlab{a}}, \apj, 689, 127

\bibitem[{{Pope} {et~al.}(2008{\natexlab{b}}){Pope}, {Chary}, {Alexander},
  {Armus}, {Dickinson}, {Elbaz}, {Frayer}, {Scott}, \& {Teplitz}}]{pope08smg}
---. 2008{\natexlab{b}}, \apj, 675, 1171

\bibitem[{{Pope} {et~al.}(2006){Pope}, {Scott}, {Dickinson}, {Chary},
  {Morrison}, {Borys}, {Sajina}, {Alexander}, {Daddi}, {Frayer}, {MacDonald},
  \& {Stern}}]{pope06}
---. 2006, \mnras, 370, 1185

\bibitem[{{Ptak} {et~al.}(2003){Ptak}, {Heckman}, {Levenson}, {Weaver}, \&
  {Strickland}}]{ptak03}
{Ptak}, A., {Heckman}, T., {Levenson}, N.~A., {Weaver}, K., \& {Strickland}, D.
  2003, \apj, 592, 782

\bibitem[{{Richards}(2000)}]{richards00}
{Richards}, E.~A. 2000, \apj, 533, 611

\bibitem[{{Rieke} {et~al.}(2009){Rieke}, {Alonso-Herrero}, {Weiner},
  {P{\'e}rez-Gonz{\'a}lez}, {Blaylock}, {Donley}, \& {Marcillac}}]{rieke09}
{Rieke}, G.~H., {Alonso-Herrero}, A., {Weiner}, B.~J.,
  {P{\'e}rez-Gonz{\'a}lez}, P.~G., {Blaylock}, M., {Donley}, J.~L., \&
  {Marcillac}, D. 2009, \apj, 692, 556

\bibitem[{{Rigby} {et~al.}(2008){Rigby}, {Marcillac}, {Egami}, {Rieke},
  {Richard}, {Kneib}, {Fadda}, {Willmer}, {Borys}, {van der Werf},
  {P{\'e}rez-Gonz{\'a}lez}, {Knudsen}, \& {Papovich}}]{rigby08}
{Rigby}, J.~R., {et~al.} 2008, \apj, 675, 262

\bibitem[{{Sajina} {et~al.}(2007){Sajina}, {Yan}, {Armus}, {Choi}, {Fadda},
  {Helou}, \& {Spoon}}]{sajina07}
{Sajina}, A., {Yan}, L., {Armus}, L., {Choi}, P., {Fadda}, D., {Helou}, G., \&
  {Spoon}, H. 2007, \apj, 664, 713

\bibitem[{{Sanders} {et~al.}(1988){Sanders}, {Soifer}, {Elias}, {Madore},
  {Matthews}, {Neugebauer}, \& {Scoville}}]{sanders88}
{Sanders}, D.~B., {Soifer}, B.~T., {Elias}, J.~H., {Madore}, B.~F., {Matthews},
  K., {Neugebauer}, G., \& {Scoville}, N.~Z. 1988, \apj, 325, 74

\bibitem[{{Shi} {et~al.}(2009){Shi}, {Rieke}, {Lotz}, \&
  {Perez-Gonzalez}}]{shi09}
{Shi}, Y., {Rieke}, G., {Lotz}, J., \& {Perez-Gonzalez}, P.~G. 2009, \apj, 697,
  1764

\bibitem[{{Shi} {et~al.}(2006){Shi}, {Rieke}, {Hines}, {Gorjian}, {Werner},
  {Cleary}, {Low}, {Smith}, \& {Bouwman}}]{shi06}
{Shi}, Y., {et~al.} 2006, \apj, 653, 127

\bibitem[{{Sturm} {et~al.}(2006){Sturm}, {Hasinger}, {Lehmann}, {Mainieri},
  {Genzel}, {Lehnert}, {Lutz}, \& {Tacconi}}]{sturm06}
{Sturm}, E., {Hasinger}, G., {Lehmann}, I., {Mainieri}, V., {Genzel}, R.,
  {Lehnert}, M.~D., {Lutz}, D., \& {Tacconi}, L.~J. 2006, \apj, 642, 81

\bibitem[{{Sturm} {et~al.}(2005){Sturm}, {Schweitzer}, {Lutz}, {Contursi},
  {Genzel}, {Lehnert}, {Tacconi}, {Veilleux}, {Rupke}, {Kim}, {Sternberg},
  {Maoz}, {Lord}, {Mazzarella}, \& {Sanders}}]{sturm05}
{Sturm}, E., {et~al.} 2005, \apjl, 629, L21

\bibitem[{{Swinbank} {et~al.}(2004){Swinbank}, {Smail}, {Chapman}, {Blain},
  {Ivison}, \& {Keel}}]{swinbank04}
{Swinbank}, A.~M., {Smail}, I., {Chapman}, S.~C., {Blain}, A.~W., {Ivison},
  R.~J., \& {Keel}, W.~C. 2004, \apj, 617, 64

\bibitem[{{Swinbank} {et~al.}(2010){Swinbank}, {Smail}, {Chapman}, {Borys},
  {Alexander}, {Blain}, {Conselice}, {Hainline}, \& {Ivison}}]{swinbank10}
{Swinbank}, A.~M., {et~al.} 2010, \mnras, 412

\bibitem[{{Szokoly} {et~al.}(2004){Szokoly}, {Bergeron}, {Hasinger}, {Lehmann},
  {Kewley}, {Mainieri}, {Nonino}, {Rosati}, {Giacconi}, {Gilli}, {Gilmozzi},
  {Norman}, {Romaniello}, {Schreier}, {Tozzi}, {Wang}, {Zheng}, \&
  {Zirm}}]{szokoly04}
{Szokoly}, G.~P., {et~al.} 2004, \apjs, 155, 271

\bibitem[{{Teng} {et~al.}(2005){Teng}, {Wilson}, {Veilleux}, {Young},
  {Sanders}, \& {Nagar}}]{teng05}
{Teng}, S.~H., {Wilson}, A.~S., {Veilleux}, S., {Young}, A.~J., {Sanders},
  D.~B., \& {Nagar}, N.~M. 2005, \apj, 633, 664

\bibitem[{{Treister} {et~al.}(2009){Treister}, {Cardamone}, {Schawinski},
  {Urry}, {Gawiser}, {Virani}, {Lira}, {Kartaltepe}, {Damen}, {Taylor}, {Le
  Floc'h}, {Justham}, \& {Koekemoer}}]{treister09}
{Treister}, E., {et~al.} 2009, \apj, 706, 535

\bibitem[{{Tyler} {et~al.}(2009){Tyler}, {Floc'h}, {Rieke}, {Dey}, {Desai},
  {Brand}, {Borys}, {Jannuzi}, {Armus}, {Dole}, {Papovich}, {Brown},
  {Blaylock}, {Higdon}, {Higdon}, {Charmandaris}, {Ashby}, \&
  {Smith}}]{tyler09}
{Tyler}, K.~D., {et~al.} 2009, \apj, 691, 1846

\bibitem[{{Vandame} {et~al.}(2001){Vandame}, {Olsen}, {Jorgensen},
  {Groenewegen}, {Schirmer}, {Arnouts}, {Benoist}, {da Costa}, {Mignani},
  {Rite'}, {Slijkhuis}, {Hatziminaoglou}, {Hook}, {Madejsky}, \&
  {Wicenec}}]{vandame01}
{Vandame}, B., {et~al.} 2001, ArXiv Astrophysics e-prints

\bibitem[{{Veilleux} {et~al.}(2009){Veilleux}, {Rupke}, {Kim}, {Genzel},
  {Sturm}, {Lutz}, {Contursi}, {Schweitzer}, {Tacconi}, {Netzer}, {Sternberg},
  {Mihos}, {Baker}, {Mazzarella}, {Lord}, {Sanders}, {Stockton}, {Joseph}, \&
  {Barnes}}]{veilleux09}
{Veilleux}, S., {et~al.} 2009, \apjs, 182, 628

\bibitem[{{Villar} {et~al.}(2008){Villar}, {Gallego}, {P{\'e}rez-Gonz{\'a}lez},
  {Pascual}, {Noeske}, {Koo}, {Barro}, \& {Zamorano}}]{villar08}
{Villar}, V., {Gallego}, J., {P{\'e}rez-Gonz{\'a}lez}, P.~G., {Pascual}, S.,
  {Noeske}, K., {Koo}, D.~C., {Barro}, G., \& {Zamorano}, J. 2008, \apj, 677,
  169

\bibitem[{{Weedman} {et~al.}(2006{\natexlab{a}}){Weedman}, {Le Floc'h},
  {Higdon}, {Higdon}, \& {Houck}}]{weedman06red}
{Weedman}, D.~W., {Le Floc'h}, E., {Higdon}, S.~J.~U., {Higdon}, J.~L., \&
  {Houck}, J.~R. 2006{\natexlab{a}}, \apj, 638, 613

\bibitem[{{Weedman} {et~al.}(2006{\natexlab{b}}){Weedman}, {Soifer}, {Hao},
  {Higdon}, {Higdon}, {Houck}, {Le Floc'h}, {Brown}, {Dey}, {Jannuzi}, {Rieke},
  {Desai}, {Bian}, {Thompson}, {Armus}, {Teplitz}, {Eisenhardt}, \&
  {Willner}}]{weedman06irs}
{Weedman}, D.~W., {et~al.} 2006{\natexlab{b}}, \apj, 651, 101

\bibitem[{{Werner} {et~al.}(2004){Werner}, {Roellig}, {Low}, {Rieke}, {Rieke},
  {Hoffmann}, {Young}, {Houck}, {Brandl}, {Fazio}, {Hora}, {Gehrz}, {Helou},
  {Soifer}, {Stauffer}, {Keene}, {Eisenhardt}, {Gallagher}, {Gautier}, {Irace},
  {Lawrence}, {Simmons}, {Van Cleve}, {Jura}, {Wright}, \&
  {Cruikshank}}]{werner04}
{Werner}, M.~W., {et~al.} 2004, \apjs, 154, 1

\bibitem[{{Wolf} {et~al.}(2004){Wolf}, {Meisenheimer}, {Kleinheinrich},
  {Borch}, {Dye}, {Gray}, {Wisotzki}, {Bell}, {Rix}, {Cimatti}, {Hasinger}, \&
  {Szokoly}}]{wolf04}
{Wolf}, C., {et~al.} 2004, \aap, 421, 913

\bibitem[{{Yan} {et~al.}(2005){Yan}, {Chary}, {Armus}, {Teplitz}, {Helou},
  {Frayer}, {Fadda}, {Surace}, \& {Choi}}]{yan05}
{Yan}, L., {et~al.} 2005, \apj, 628, 604

\end{thebibliography}
